\documentclass[twocolumn,usenatbib]{mnras}

\let\oldhref\href \renewcommand{\href}[2]{\oldhref{#1}{\hbox{#2}}}

\usepackage[utf8]{inputenc}

\usepackage{pex-macros}
\graphicspath{{./}{./figs/}}

\newcommand*{\inst}[1]{\ensuremath{^{#1}}}

\title[Multi-spectral planet extraction]{PeX 1.\ Multi-spectral expansion of
residual speckles for planet detection}

\author[Nicholas Devaney \& Éric Thiébaut]{Nicholas Devaney\inst{1}\thanks{NUI,
Galway} \& Éric Thiébaut\inst{2}\thanks{Univ Lyon, Univ Lyon1, Ens de Lyon, CNRS, Centre de Recherche Astrophysique de Lyon UMR5574, F-69230, Saint-Genis-Laval, France.}}

\newcommand*{\PeX}{{\sc PeX}\xspace}

\newcommand*{\RefPos}{\V{s}}
\newcommand*{\ImgPos}{\V{r}}
\newcommand*{\PupPos}{\V{u}}
\newcommand*{\PupAmp}{a}
\newcommand*{\ImgAmp}{\hat{a}}
\newcommand*{\Phase}{\phi}
\newcommand*{\FTPhase}{\hat{\Phase}}
\newcommand*{\OPD}{d}
\newcommand*{\FocalLength}{f}
\newcommand*{\StarPSF}{P}
\newcommand*{\PSFmode}{p}
\newcommand*{\SpctWgt}{g}
\newcommand*{\StarRefPSF}{\StarPSF_{\!\Tag{ref}}}
\newcommand*{\Img}{I}
\newcommand*{\InterpImg}{A}
\newcommand*{\StarSED}{F}
\newcommand*{\PlanetSED}{f}
\newcommand*{\FirstTermSED}{q} 

\begin{document}

\maketitle

\begin{abstract}
The detection of exoplanets in coronographic images is severely limited by
residual starlight speckles. Dedicated post-processing can drastically
reduce this ``stellar leakage'' and thereby increase the faintness of
detectable exoplanets.  Based on a multi-spectral series expansion of the
diffraction pattern, we derive a multi-mode model of the residuals which
can be exploited to estimate and thus remove the residual speckles in
multi-spectral coronographic images.  Compared to other multi-spectral
processing methods, our model is physically grounded and is suitable for
use in an (optimal) inverse approach.  We demonstrate the ability of our
model to correctly estimate the speckles in simulated  data and demonstrate
that very high contrasts can be achieved.  We further apply our method to
removing speckles from a real data cube obtained with the SPHERE IFS
instrument.
\end{abstract}


\section{Introduction}

Direct detection of extra-solar planet images or spectra is extremely
challenging because of the small angular separation and the enormous contrast
ratio between the parent star and the planet.  For ground-based detection of
exoplanets it is necessary to employ a very high order adaptive optics (AO)
system to reduce atmospheric wavefront errors to a very low level. In addition,
a coronograph is employed to mask light from the parent star
\citep{Mawet_et_al-2012}.  Specialized instruments have recently been
commissioned to search for exoplanets; \noun{Sphere} in the case of the 8m VLT
\citep{Beuzit_et_al_2008, Vigan_et_al_2015}, GPI in the case of the 8m Gemini
telescope \citep{Macintosh_et_al-2014,Kalas_et_al_2015}, SCExAO for the 8m
Subaru telescope \citep{Guyon_et_al-2011SPIE}, and the Project 1640 system at
the Palomar observatory \citep{Crepp_et_al-2011ApJ}.

Some exoplanets have already been detected by direct imaging
 \citep{Marois_et_al-2008, Lagrange_et_al-2009, Lafreniere_et_al-2010,
Ireland_et_al-2011,Rameau_et_al-2013, Bonavita_et_al-2014}, although the
separations of the planets from the parent stars are relatively large.  It has
been found that the detection of exoplanets in AO-corrected images is severely
limited by the presence of residual speckles
\citep[\eg,][]{Janson_et_al-2006-GQLupi}. These speckles arise from uncorrected
atmospheric errors, and uncorrected optical errors in the telescope and/or
imaging instrument.  The residual atmospheric errors are random and will
average to a smooth halo if the exposure time is long compared with the
atmospheric coherence time.  However, the speckles due to uncorrected optical
errors evolve slowly with time \citep{Hinkley_et_al-2007} and are difficult to
distinguish from real point sources in the image.  In order to reduce these
static speckles, a number of differential imaging techniques have been
proposed.

In angular differential imaging (ADI), the field is allowed to rotate during
the observation (\eg, on an altitude-azimuth telescope) while the instrument
and AO system are fixed in position; it is supposed that the system point
spread function (PSF), including the quasi-static speckles, will remain
approximately constant during the observation, while any exoplanets will rotate
about the central star.  In the LOCI approach which was first proposed by
\citet{Lafreniere_et_al-2007ApJ}, the image is divided into annular regions
defined by annuli and wedges, and the PSF is estimated and subtracted in each
region. A temporal weighting of the frames making up the local PSF estimate can
therefore vary over the image, reflecting the fact that the temporal evolution
of speckles depends on their distance from the center of the PSF.  The direct
application of LOCI to IFS data can lead to errors in the exoplanet
spectro-photometry due to subtraction of residual starlight.
\citet{Pueyo_et_al-2012-damped_LOCI} proposed a modified algorithm referred to
as ``\emph{Damped LOCI}'' in which the cost function is modified to attempt to
conserve companion flux. \citet{Marois_et_al-2014} show that the performance
can be improved by including a prior model of the planet spectrum --- this is
referred to as template-LOCI or simply TLOCI.

An alternative approach, sometimes referred to as \emph{KLIP}
(Karhunen-Lo\`{e}ve Image Projection), to analysing ADI data is based on
carrying out a Principal Component Analysis (PCA) of the sequence of images
\citep{Soummer_et_al_2012,Amara_Quanz-2012}.  The images are concatenated into
a large array, and a singular value decomposition carried out. This identifies
orthogonal modes of variation. Each image is subsequently decomposed onto a
number of principal components and the result is subtracted from the original
images. The resulting residues are de-rotated and summed, following which any
planet candidates are identified by thresholding. The number of modes to use is
the result of a compromise between complete subtraction of the stellar signal
and the removal of planet flux. The trade-off is usually carried out by
injecting synthetic planets into the data and determining which number of modes
will maximise the planet signal to noise ratio. In addition, the optimal number
of modes will depend on the radial distance from the star at which it is
required to search for a planet \citep{Meshkat_et_al-2013}.

Another type of differential imaging relies on obtaining images at more than
one wavelength.  Spectral Differential Imaging (SDI) \citep{Smith-1987,
  Racine_et_al-1999} uses images obtained (preferably simultaneously) at
different wavelengths. If one of the images is taken inside a Methane
absorption line at $1.6\,\micron$ then it can be used as a good PSF estimate
since the signal from any exoplanet would be much weaker, at least if the
planet is a gas giant.  \citet{Marois_et_al-2000} developed implementation
details for the idea and this approach has been implemented in instruments used
to survey stars for methane-rich companions \citep{Biller_07,
  Marois_et_al-2005}.  While the technique may be extended to include more
wavelengths using double and higher order differencing, it is not
straightforward and, in practice, is limited by noise
\citep{Marois_et_al-2005}.

Most of the instruments to be used for exoplanet detection will include an
Integral Field Spectrograph (IFS) in order to provide spectral information on
any detected planet. It has been realized that the information present in IFS
data may be used to reduce the effect of the stellar PSF, including any static
speckles.  \citet{Thatte_et_al-2007} describe a technique in which the IFS data
cube is radially rescaled to remove the wavelength-dependent scaling of the
diffraction pattern. The PSF is then estimated by polynomial fitting along the
wavelength axis for each pixel in the rescaled data cube. The polynomial order
of the fit is chosen by the user --- a higher order will tend to remove any
planet signal and, in practice, a low order is used. Subtraction of the PSF
estimate will allow detection of faint companions. This approach, which was
originally proposed by \citet{Sparks_Ford-2002-imaging_spectroscopy} is
referred to as Spectral Deconvolution (\emph{SD}).  \citet{Crepp_et_al-2011ApJ}
describe a modification of the LOCI algorithm for use with IFS data cubes.  A
PSF reference is built up for each wavelength channel by combining images which
are nearby in both wavelength and time (but separated enough in wavelength to
ensure the speckles have moved significantly with respect to the diffraction
limit). The frames at different wavelengths are scaled and combined by means of
least squares.  Using data from the Palomar observatory system, they find a
small departure from linear scaling with wavelength, and attribute this to
out-of-pupil aberrations. The Principal Component Analysis technique can also
be applied to IFS data.

Here we propose a new approach to the detection of exoplanets in multi-spectral
data.  \citet{Perrin_et_al-2003-PSF} proposed that the PSF can be expanded in a
power series of spatial modes which are functions of the residual phase in the
pupil. The modes of odd order are antisymmetric, while the even order modes are
symmetric. The dominant modes depend on the Strehl ratio of the corrected
wavefront and the distance from the center of the PSF. We extend this work to
consider the spectral dependence of these modes, which turns out to be power
laws. We examine the validity of this analysis using a Singular Value
Decomposition of simulated multi-spectral coronographic images, and show that
using this approach can give excellent speckle suppression in simulated data.
The approach has similarities with PCA, but there are some important
differences.  In particular, our method is based on imaging physics (the power
series expansion of the PSF), which leads to insights not available with ad-hoc
processing. We will show that is it more flexible than approaches based on
Singular Value Decomposition, including PCA. Finally, it lends itself to an
inverse approach for the joint estimation of the residual PSF plus detection of
planets. A preliminary version of this work has been presented
\citep{Thiebaut_et_al-2016-physical_constraints}. In this paper we will provide more
details about the derivation and the testing of our model and show how to use
this model to perform planet detection in multi-spectral and multi-temporal
data. We present an example of planet detection using data from the \noun{Sphere} IFS,  but the
full exploitation of our technique in an inverse approach will be presented in
a subsequent paper.

\section{Model of the stellar leakage}

In order to achieve the best detection sensitivity it is necessary to reduce
residual speckles as much as possible by appropriate post-processing of the
images.  We propose to tackle the removal of the stellar leakage as an
\emph{inverse problem} based on a proper modeling of the on-axis PSF.  In order
to constrain this model and thus achieve a very good level of ``\emph{soft
coronography}'', we exploit the chromatic behavior of the speckle pattern.

\subsection{Speckle Alignment}

As described in the introduction, most speckle removal methods work on image
differences where the observed images are obtained at different times and/or
wavelengths and are subtracted after an interpolation and a multiplication by a
scaling factor of one of the images.  The interpolation implements a
geometrical transformation of coordinates intended to align the speckles in the
two images.  Typically, this geometrical transform accounts for
translation (to compensate for pointing errors and jitter), rotation (to
compensate for pupil rotation during the night) and magnification (to
compensate for chromatic geometrical effects due to diffraction).  In
principle, it is also possible to compensate for other geometrical effects such
as distortion.  The scaling factor accounts for any variation of the flux
received from the star as it is seen in the two images (\eg due to the
variation of transparency or to the star spectrum if the two images are at
different wavelengths).  The resulting image differences are called ADI
(angular differential images) or SDI (spectral differential images) depending
on whether the two images  come from different exposures (and hence have different
rotation angles) or from different spectral channels. The idea is that the
residual speckles are cancelled or at least strongly reduced in the differential
images compared to the original data.  In principle, it is possible to combine
images at different times \emph{and} different wavelengths.

Since we want to first focus on speckle removal, we start by considering the
case of observing an isolated star with no surrounding sources.   In this case,
the light distribution in the focal plane of a telescope is simply given by:
\begin{equation}
  \label{eq:star-image}
  \Img(\ImgPos,\lambda,t)
  = \StarSED(\lambda) \, \StarPSF(\ImgPos,\lambda,t)
\end{equation}
where $\ImgPos$ is the position in the focal plane, $\lambda$ is the
wavelength, $t$ is the time, $\StarSED(\lambda)$ is the spectral energy
distribution (SED) of the star and $\StarPSF(\ImgPos,\lambda,t)$ is the
on-axis\footnote{What we call the \emph{on-axis direction} is determined by the
direction toward the star not that of the optical axis even though they usually
coincide.} point spread function (PSF).  Introducing 2D spatial coordinates
$\RefPos$ in a reference coordinate system where the speckles are aligned, we
can rewrite the on-axis PSF as:
\begin{equation}
  \label{eq:on-axis-ref-psf}
  \StarPSF(\ImgPos,\lambda,t)
  = \StarRefPSF\Paren{\RefPos,\lambda,t}
  \, \Abs*{\frac{\partial \RefPos}{\partial \ImgPos}} \, ,
\end{equation}
where $\StarRefPSF(\RefPos,\lambda,t)$ is the distribution of speckles in the
reference coordinate system and $\Abs{\partial \RefPos/\partial \ImgPos}$ is
the absolute value of the determinant of the Jacobian matrix of the coordinate
transform $\ImgPos \mapsto \RefPos$ and is needed to insure proper
normalization.  If there are no losses, the distributions of speckles are both
normalized in their respective coordinate systems:
\begin{equation}
  \label{eq:PSF-normalization}
  \iint \StarPSF(\ImgPos,\lambda,t) \, \mathd^2 \ImgPos
  = \iint \StarRefPSF(\RefPos,\lambda,t) \, \mathd^2 \RefPos = 1 \, .
\end{equation}
This normalization condition insures that:
\begin{equation}
  \label{eq:image-normalization}
  \iint \Img(\ImgPos,\lambda,t) \, \mathd^2 \ImgPos = \StarSED(\lambda) \, .
\end{equation}
In these equations, the term $\StarSED(\lambda)$ is the SED of the star
\emph{as seen by the instrument}, \ie it takes into account the transmission by
the atmosphere and the instrument which may depend on wavelength.  The star SED
could be written as $\StarSED(\lambda,t)$, if the transmission also depends on
time $t$.

The distribution $\StarRefPSF(\RefPos,\lambda,t)$ can also be seen as the
point spread function (PSF) in the reference coordinate system for a source in
the direction of the star.  We will refer to it as the ``\emph{reference
on-axis PSF}''. The mapping $\ImgPos \leftrightarrow \RefPos$ is a general
formalization of the geometrical coordinate transform implemented by existing
methods such as ADI or SDI. The key idea of ADI- or SDI-based methods is that
the reference on-axis PSF is approximately independent of the wavelength
and/or the time and can thus be canceled by means of image subtraction after
interpolation.

\subsection{Spatio-spectral distribution of the speckles}
\label{sec:diffraction}

Following the work of \citet{Perrin_et_al-2003-PSF} who expanded the PSF as an
infinite Taylor series with respect to the phase aberration, we propose to
derive a model for the distribution of speckles which accounts for
chromatic effects.  Our intention is to use this model to achieve a better
suppression of the stellar speckles.

Since the source is effectively at an infinite distance, the illumination in the
image plane results from Fraunhofer diffraction and the on-axis PSF is given
by:
\begin{equation}
  \label{eq:diffraction}
  \StarPSF(\ImgPos,\lambda,t)
  = \frac{1}{\rho(\lambda)} \, \Abs*{
    \iint \!\! \PupAmp(\PupPos,\lambda,t) \,
    \mathe^{
      \frac{\mathi\,2\,\pi}{\FocalLength\,\lambda} \,
      \Inner{\PupPos}{\ImgPos}
    } \, \mathd^2\PupPos
  }^2
\end{equation}
with $\rho(\lambda)$ a normalization factor, $\PupAmp(\PupPos,\lambda,t)$
the complex amplitude transmission at the position $\PupPos$ of the pupil
plane and $\FocalLength$ the focal length.  The expression
$\Inner{\PupPos}{\ImgPos}$ denotes the usual scalar product of $\PupPos$ by
$\ImgPos$.  In the conditions considered here, the distribution of the
diffracted light depends on the phase aberrations in the pupil plane.
Propagation of aberrations arising away from the pupil plane can give rise
to a chromatic effect \citep{Marois_et_al-2006} which is not included in
this analysis.

Also neglecting the chromaticism of the refractive index of air, the complex
amplitude transmission is given by:
\begin{equation}
  \label{eq:complex-transmission}
  \PupAmp(\PupPos,\lambda,t) =
    \PupAmp_0(\PupPos) \,
    \exp\Paren*{\frac{\mathi\,2\,\pi}{\lambda}\,\OPD(\PupPos,t)} \, ,
\end{equation}
where $\PupAmp_0(\PupPos)$ is the aberration-free telescope complex amplitude
transmission (the so-called \emph{pupil function}) and $\OPD(\PupPos,t)$ is an
achromatic optical path difference due to the aberrations.  The integral in
Eq.~(\ref{eq:diffraction}) is directly related to the spatial (inverse) Fourier
transform of the complex amplitude transmitted by the pupil:
\begin{equation}
  \ImgAmp(\V{\omega},\lambda,t)
  = \iint \PupAmp(\PupPos,\lambda,t) \,
    \mathe^{\mathi\,2\,\pi \, \Inner{\PupPos}{\V{\omega}}}
    \mathd^2\PupPos \, ,
\end{equation}
which is proportional to the complex amplitude in the focal plane.
Introducing this quantity in Eq.~(\ref{eq:diffraction}) yields:
\begin{equation}
  \label{eq:diffraction2}
  \StarPSF(\ImgPos,\lambda,t)
  = \frac{1}{\rho(\lambda)} \, \Abs*{
    \ImgAmp\Paren*{\frac{\ImgPos}{\FocalLength\,\lambda},\lambda,t}
  }^2 \, .
\end{equation}
The factor $\rho(\lambda)$ is such that the normalization in
Eq.~(\ref{eq:PSF-normalization}) holds and therefore:
\begin{align}
  \rho(\lambda)
  &= \iint
\Abs*{\ImgAmp\Paren*{\frac{\ImgPos}{\FocalLength\,\lambda},\lambda,t}}^2
    \, \mathd^2\ImgPos
   \notag \\
  &= (\FocalLength\,\lambda)^2 \,
  \iint \Abs*{\ImgAmp\Paren*{\V{\omega},\lambda,t}}^2 \, \mathd^2\V{\omega}
   \notag \\
  &= (\FocalLength\,\lambda)^2 \,
  \iint \Abs*{\PupAmp_0\Paren*{\PupPos}}^2 \, \mathd^2\PupPos \, ,
  \label{eq:normalization-factor}
\end{align}
with $\V{\omega} = \ImgPos/(\FocalLength\,\lambda)$ and where the latter
equation follows from Parseval's theorem and from
Eq.~(\ref{eq:complex-transmission}).

Following \citet{Perrin_et_al-2003-PSF}, the exponential term of the complex
amplitude transmitted by the pupil can be expanded in an absolutely convergent
series in the optical path difference $\OPD(\PupPos,t)$:
\begin{displaymath}
  \PupAmp(\PupPos,\lambda,t)
  = \PupAmp_0(\PupPos) \,
    \sum_{k \ge 0} \frac{1}{k!}\,\Paren*{
      \frac{\mathi\,2\,\pi}{\lambda}\,\OPD(\PupPos,t)
    }^{\!\!k}
    \, .
\end{displaymath}
At this point it is convenient to introduce the phase aberration in the pupil
at a given reference wavelength $\lambda^\Tag{ref}$:
\begin{equation}
  \label{eq:phase-aberration}
  \Phase(\PupPos,t)
  = \frac{2\,\pi}{\lambda^\Tag{ref}}\,\OPD(\PupPos,t)
    \, ,
\end{equation}
and to rewrite the expanded complex amplitude as:
\begin{equation}
  \label{eq:complex-transmission-expansion}
  \PupAmp(\PupPos,\lambda,t)
  = \PupAmp_0(\PupPos) \,
    \sum_{k \ge 0}
    \frac{\mathi^k\,\gamma(\lambda)^k\,\Phase(\PupPos,t)^{k}}{k!}
    \, ,
\end{equation}
with:
\begin{equation}
  \label{eq:chromatic-scaling}
  \gamma(\lambda) = \frac{\lambda^\Tag{ref}}{\lambda} \, .
\end{equation}
Then taking the inverse Fourier transform of
Eq.~(\ref{eq:complex-transmission-expansion}) yields:
\begin{equation}
  \label{eq:image-amplitude-expansion}
  \ImgAmp(\V{\omega},\lambda,t)
  = \sum_{k \ge 0} \frac{
      \mathi^k \, \gamma(\lambda)^k \, \xi_k(\V{\omega},t)
    }{
      k!
    } \, .
\end{equation}
where:
\begin{align}
  \xi_k(\V{\omega},t)
  &= \iint \PupAmp_0(\PupPos) \, \Phase(\PupPos,t)^{k} \,
      \mathe^{\mathi\,2\,\pi \, \Inner{\PupPos}{\V{\omega}}}
      \mathd^2\PupPos
      \notag \\
  &= \ImgAmp_0(\V{\omega}) \star^k \FTPhase(\V{\omega},t) \,
  \label{eq:img-amp-basis}
\end{align}
with $\ImgAmp_0(\V{\omega})$ and $\FTPhase(\V{\omega},t)$ the inverse spatial
Fourier transforms of the aberration-free pupil transmission and phase
aberration at the reference wavelength.  As in \citet{Perrin_et_al-2003-PSF},
we use $\star^k$ to denote multiple convolution products over the conjugate
position $\V{\omega}$:
\begin{align}
  \ImgAmp_0\star^{k}\FTPhase
  \bydef \left\{
    \begin{array}{ll}
      \ImgAmp_0 & \text{if $k = 0$\,,} \\
      \displaystyle\ImgAmp_0\star
      \overbrace{
        \FTPhase\star\ldots\star\FTPhase
      }^{\text{$k$ terms}} & \text{if $k > 0$\,,} \\
    \end{array}
    \right.
\end{align}
with $\star$ the ordinary convolution product.  Taking the squared modulus of
the complex amplitude in the focal plane and grouping the terms of the same
order with respect to the phase aberrations yields:
\begin{align}
  \Abs{\ImgAmp(\V{\omega},\lambda,t)}^2
  &=
  \sum_{k_1 \ge 0}\sum_{k_2 \ge 0}
  \frac{
    \mathi^{k_1 - k_2} \, \gamma(\lambda)^{k_1 + k_2} \,
    \xi_{k_1}(\V{\omega},t) \,
    \xi_{k_2}^{*}(\V{\omega},t)
  }{
    {k_1}!\,{k_2}!
  } \notag \\
  &= \sum_{k \ge 0} \gamma(\lambda)^{k} \, (-\mathi)^{k}
     \sum_{k' = 0}^{k} \frac{(-1)^{k'} \, \xi_{k'}(\V{\omega},t) \,
    \xi_{k-k'}^{*}(\V{\omega},t)}{k'! \, (k - k')!} \, . \notag
\end{align}
Taking the normalization in Eq.~(\ref{eq:normalization-factor}) into account,
the on-axis PSF can finally be written:
\begin{equation}
  \label{eq:psf-expansion}
  \boxed{
    \StarPSF(\ImgPos,\lambda,t) = \sum_{k \ge 0} \gamma(\lambda)^{k + 2} \,
    \PSFmode_k\Paren[\big]{\gamma(\lambda)\,\ImgPos,t} \, ,
  }
\end{equation}
with $\gamma(\lambda) = \lambda^\Tag{ref}\!/\lambda$ and where the
\emph{on-axis PSF modes} $\PSFmode_k(\RefPos,t)$, with $\RefPos =
\gamma(\lambda)\,\ImgPos$, are given by:
\begin{equation}
  \label{eq:psf-modes}
  \PSFmode_k(\RefPos,t) = \frac{(-\mathi)^{k}}{\rho(\lambda^\Tag{ref})} \,
    \sum_{k' = 0}^{k} \frac{
      (-1)^{k'} \,
      \xi_{k'}\Paren*{\V{\omega},t} \,
      \xi_{k-k'}^{*}\Paren*{\V{\omega},t}
    }{
      k'! \, (k - k')!
    } \, ,
\end{equation}
where:
\begin{displaymath}
  \V{\omega}
  = \frac{\RefPos}{\FocalLength\,\lambda^\Tag{ref}}
  = \frac{\gamma(\lambda)}{\FocalLength\,\lambda^\Tag{ref}} \, \ImgPos\,.
\end{displaymath}
In words, the expansion in Eq.~(\ref{eq:psf-expansion}) shows that the
change of the PSF with wavelength is a combination of chromatic
magnification, by the factor $\gamma(\lambda)$, and amplifications, by
powers of $\gamma(\lambda)$. The important point is that there are no other
wavelength dependencies in the on-axis PSF.  In particular, the PSF modes
$\PSFmode_k(\RefPos,t)$ are achromatic; they only depend on the position
$\RefPos$ in the reference coordinate system and on the time $t$.  The
first term $\PSFmode_0$ of the PSF expansion is the PSF without aberrations
at the reference wavelength, while the other terms are due to the phase
aberrations.

Our equations extend the work by \citet{Perrin_et_al-2003-PSF} who considered
the monochromatic case. In particular they did not consider the chromatic
magnification and amplification of the PSF modes.  Some interesting properties
of the on-axis PSF modes $\PSFmode_k(\RefPos,t)$ defined in
Eq.~(\ref{eq:psf-modes}) can be inferred from the paper of
\citet{Perrin_et_al-2003-PSF}.  First, all the terms $\PSFmode_k(\RefPos,t)$ of
the PSF expansion are real-valued, the zeroth order term
$\PSFmode_0(\RefPos,t)$ of this series is the unaberrated PSF, and all the odd
terms are spatially antisymmetric while all the even terms are symmetric. Thus,
for $k > 0$, the terms $\PSFmode_k(\RefPos,t)$ have the same parity as $k$ with
respect to $\RefPos$: $\PSFmode_{2k}(-\RefPos,t) = \PSFmode_{2k}(\RefPos,t)$
while $\PSFmode_{2k+1}(-\RefPos,t) = -\PSFmode_{2k+1}(\RefPos,t)$. Second, the
model derived from the series expansion in the case of simple Fraunhofer
diffraction remains approximately valid for an apodized Lyot coronograph.  We
therefore expect that the on-axis PSF model in Eq.~(\ref{eq:psf-expansion}) can
serve as a good basis to remove the stellar leakage in the proposed inverse
approach. Which terms of the expansion dominate depends on the Strehl ratio,
distance from the axis and on the attenuation by the coronograph.

\subsection{On-axis PSF in the reference coordinate system}
\label{sec:ref-on-axis-PSF}

If one simply defines the position in the reference coordinate system as:
\begin{equation}
  \label{eq:simple-mapping}
  \RefPos = \gamma(\lambda) \, \ImgPos \, ,
\end{equation}
then noting that it yields:
\begin{equation}
  \Abs{\partial\RefPos/\partial\ImgPos} = \gamma(\lambda)^{2} \, ,
\end{equation}
and combining Eq.~(\ref{eq:on-axis-ref-psf}) and Eq.~(\ref{eq:psf-expansion}),
the reference on-axis PSF is written:
\begin{equation}
  \label{eq:ref-psf-expansion}
  \boxed{
  \StarRefPSF(\RefPos,\lambda,t)
  = \sum_{k \ge 0} \gamma(\lambda)^{k} \, \PSFmode_k\Paren{\RefPos,t} \, ,
  }
\end{equation}
with $\gamma(\lambda) = \lambda^\Tag{ref}\!/\lambda$.  Since there is no
chromatic spatial distortion in this expression, it is clear that the speckles
are aligned in such a reference system.

The diffraction computations in the previous section were carried out assuming
that the optical axis and the pupil orientation are the same at all
wavelengths and times. This implies that a simple chromatic magnification
between $\RefPos$, the position in the reference coordinate system and
$\ImgPos$, the position in the image coordinate system is sufficient to align
the speckles.  In practice, there may be several causes of misalignment that
must be taken into account and a more complex mapping $\RefPos \leftrightarrow
\ImgPos$ may have to be considered.  Nevertheless the series in
Eq.~(\ref{eq:ref-psf-expansion}) and relation (\ref{eq:on-axis-ref-psf})
remain valid in the general case and can be exploited to model the chromatic
behavior of the distribution of speckles.

\section{Validation of the model}
\label{sec:model-validation}

To validate our model, we propose to check how well it is able to fit realistic
simulations of the stellar leakage.  The model becomes separable if one
considers resampled images in the reference coordinate system.  Working with
resampled images, truncated singular value decomposition (TSVD) provides an
approximation which is different from our model as it makes no other
assumptions besides separability.  We use TSVD to exhibit the actual chromatic
behavior of the stellar leakage and compare it with the predictions of our
model.  The ability of our model to fit the stellar leakage is compared to TSVD
which provides the best possible fit for a given number of modes.

\subsection{Interpolated image in the reference coordinate system}
\label{sec:interpolated-image}

The PSF expansion in Eq.~(\ref{eq:psf-expansion}) yields the following
expression for the brightness distribution due to the star:
\begin{equation}
  \label{eq:star-img-expansion}
   \Img(\ImgPos,\lambda,t) = \StarSED(\lambda) \,
   \sum_{k \ge 0} \gamma(\lambda)^{2+k} \,
   \PSFmode_k\Paren[\big]{\RefPos(\ImgPos,\lambda,t),t} \, .
\end{equation}
Considering this expression, it is useful to define:
\begin{align}
  \InterpImg(\RefPos,\lambda,t)
  &\equiv \Img(\ImgPos(\RefPos,\lambda,t),\lambda,t) \notag\\
  &= \sum_{k \ge 0} \SpctWgt_k(\lambda) \, \PSFmode_k(\RefPos,t) \, ,
  \label{eq:interp-image}
\end{align}
where $\SpctWgt_k(\lambda)$ are \emph{spectral weighting functions} given by:
\begin{equation}
  \label{eq:spectral-weights}
  \SpctWgt_k(\lambda) = \StarSED(\lambda) \, \gamma(\lambda)^{2+k} \, .
\end{equation}
The quantity $\InterpImg(\RefPos,\lambda,t)$ is the observed light distribution
interpolated into the reference coordinate system.  This distribution is
clearly a separable expansion whose terms are the product of a chromatic
weight, $\SpctWgt_k(\lambda)$, by a spatio-temporal mode,
$\PSFmode_k(\RefPos,t)$.  We will see that this description is very useful as
it allows us to introduce priors on the distribution of the speckles and to
help suppress them.

The notations $\RefPos(\ImgPos,\lambda,t)$ in Eq.~(\ref{eq:star-img-expansion})
and $\ImgPos(\RefPos,\lambda,t)$ in Eq.~(\ref{eq:interp-image}) make explicit
the spatio-spectro-temporal dependency between the image coordinates $\ImgPos$
and the position $\RefPos$ in the reference coordinate system as discussed in
Section~\ref{sec:ref-on-axis-PSF}. In order to simplify notation, this
relationship will be implicitly assumed in what follows.

In practice, the data consists of a number of exposures acquired in different
spectral channels.  We denote the value measured by the $j$-th pixel in the $\ell$-th
spectral channel during the $m$-th exposure by:
\begin{equation}
  \label{eq:sampled-data}
  \Img_{j,\ell,m} \approx
  \Img(\ImgPos_{j,\ell,m},\lambda_\ell,t_m)
  \, \Delta\Omega_{j,\ell,m} \, .
\end{equation}
Here $\ImgPos_{j,\ell,m}$ is the position of the considered pixel,
$\lambda_\ell$ is the effective wavelength of the spectral channel and $t_m$ is
the mean time of the exposure.  The term $\Delta\Omega_{j,\ell,m}$ accounts for
the effective pixel area, spectral bandwidth and exposure duration.  The
$\approx$ sign in Eq.~(\ref{eq:sampled-data}) accounts for any approximations
(such as sampling the distribution instead of integrating it over
$\Delta\Omega_{j,\ell,m}$) and for the noise.

As described previously, in order to align the residual speckles, it is
necessary to resample the data into the reference coordinate system.  In
practice, this amounts to applying a linear transform to the data.  Formally:
\begin{equation}
  \InterpImg(\RefPos_i,\lambda_\ell,t'_i)
  \approx \InterpImg_{i,\ell}
  = \sum_{j,m} \, \Paren{\M{R}_{\ell,m}}_{i,j} \, \Img_{j,\ell,m} \, ,
  \label{eq:resampling}
\end{equation}
where $\M{R}_{\ell,m}$ is an interpolation operator which maps a distribution
sampled in the image frame of the $\ell$-th spectral channel during $m$-th
exposure to the reference  coordinate system.  As explained in
Section~\ref{sec:ref-on-axis-PSF}, the mapping to the reference coordinate
system depends on the wavelength and on the time, hence the corresponding
linear operator is indexed by $\ell$ and $m$.  As implicitly assumed by
Eq.~(\ref{eq:resampling}), the resampling operator $\M{R}_{\ell,m}$ also scales
the data to eliminate the $\Delta\Omega_{j,\ell,m}$ factors.  Combining
Eq.~(\ref{eq:interp-image}) and (\ref{eq:resampling}), the stellar light in the
resampled data is approximated by:
\begin{equation}
  \InterpImg_{i,\ell}
  \approx \sum_{k \ge 0}
   \PSFmode_k(\RefPos_i,t'_i) \, \SpctWgt_k(\lambda_\ell) \, .
  \label{eq:separable-model-series}
\end{equation}
In Eq.~(\ref{eq:resampling}) and~(\ref{eq:separable-model-series}), $\RefPos_i$
 and $t'_i$ involve a sampling of the position in the reference coordinate
system and of the time.  Specific time sampling (\ie not necessarily the same
as the exposures) is required to correctly account for the temporal variations
of the speckles and to ensure that the approximation assumed in
Eq.~(\ref{eq:resampling}) and~(\ref{eq:separable-model-series}) holds.

\begin{figure*}

  \centering
  \includegraphics[scale=0.45]{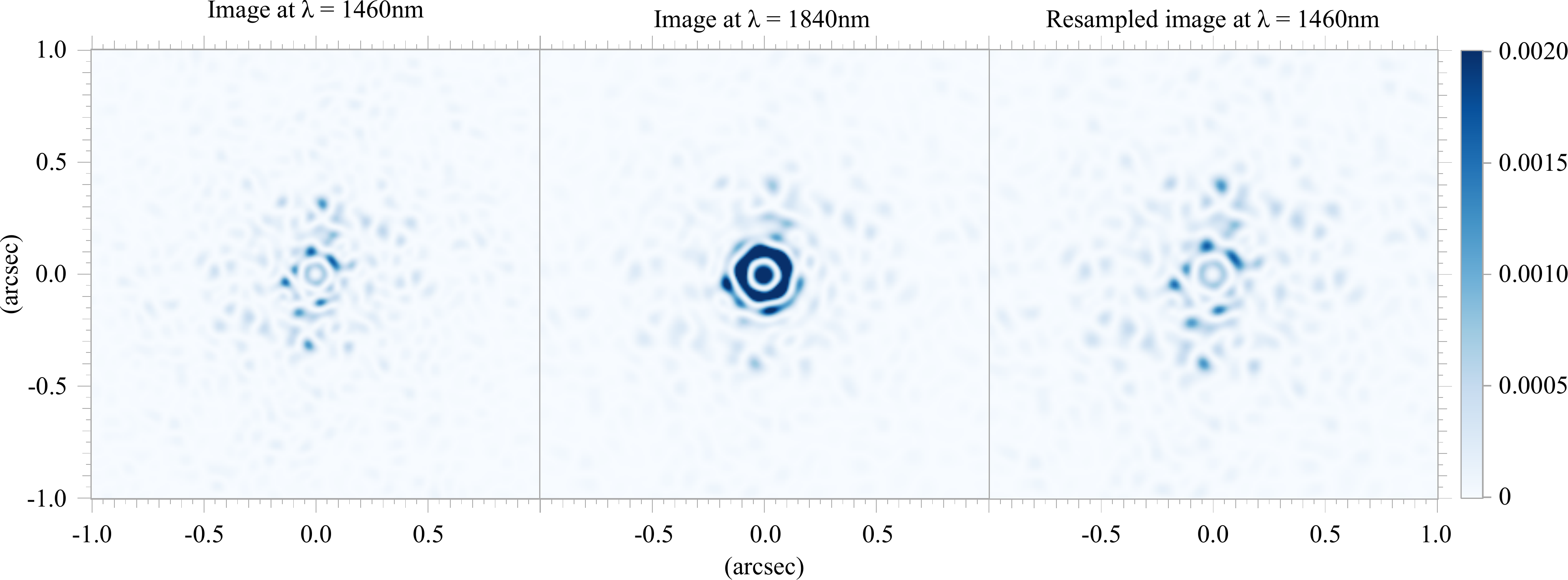}

  \caption{Simulated coronographic images.  The leftmost (resp.\ central)
  image is for the spectral channel at the shortest (resp. longest)
  wavelength.  The rightmost image is the image at the shortest wavelength
  magnified to match the diffraction pattern at the longest wavelength.
  The intensities have been normalized by the peak intensity without a
  coronograph. The simulations were carried out using parameters typical of
  \noun{Sphere} in the H band, see the text for details.
  \label{fig:images}}
\end{figure*}

\subsection{Truncated approximation}
\label{sec:truncated-approximation}

In practice, the model in Eq.~(\ref{eq:separable-model-series}) cannot be used
with an unlimited number of modes. \citet{Perrin_et_al-2003-PSF} have shown
that the higher order modes are insignificant (the series is absolutely
convergent) and that the first modes may also be
negligible in the case of images obtained using a coronograph.  Limiting the series to $K$ modes and assuming $k_0$ is the
index of the first significant term in the series, our so-called
\PeX (Planet eXtractor) model of the brightness distribution in a
reference coordinate system can be written:
\begin{equation}
  \label{eq:PeX-model1}
  \InterpImg_{i,\ell} \approx \sum_{k=1}^{K}
  \PSFmode_{k,i} \, \SpctWgt_{k,\ell}
  \quad\text{with}\quad\left\{\!\!
  \begin{array}{r@{\:}l}
  \PSFmode_{k,i} &= \PSFmode_{k + k_0 - 1}(\RefPos_i,t'_i) \\[1ex]
  \SpctWgt_{k,\ell} &=  \SpctWgt_{k + k_0 - 1}(\lambda_\ell)
  \end{array}\right.
\end{equation}
According to the results of \citet{Perrin_et_al-2003-PSF}, we anticipate that
$k_0 = 0$ without a coronograph and that $k_0 > 0$ with a coronograph.  Using
matrix notation, our approximation can be written in the more compact form:
\begin{equation}
  \M{\InterpImg}
  \approx \sum_{k=1}^{K} \V{\PSFmode}_{k} \cdot \V{\SpctWgt}_{k}\T
  \label{eq:PeX-model1-compact}
\end{equation}
where $\V{\PSFmode}_{k} = (\PSFmode_{k,1}\ \PSFmode_{k,2}\ \ldots)\T$ and
$\V{\SpctWgt}_{k} = (\SpctWgt_{k,1}\ \SpctWgt_{k,2}\ \ldots)\T$ are the
(sampled) significant PSF modes and their spectral weights.

Introducing $\StarSED_\ell = \StarSED(\lambda_\ell)$ the SED of the star in
the $\ell$-th spectral channel, and $\gamma_\ell = \gamma(\lambda_\ell) =
\lambda^\Tag{ref}\!/\lambda_\ell$, the specific form of the spectral weights
assumed by our model is given by:
\begin{equation}
  \SpctWgt_{k,\ell} = \StarSED_\ell \, \gamma_\ell^{\beta + k - 1} \, ,
  \label{eq:PeX-spectral-weights}
\end{equation}
where $\beta = k_0 + 2$ is the exponent for the first significant term of the
expansion.  With this notation, our approximation of the stellar leakage
becomes:
\begin{equation}
  \InterpImg_{i,\ell} \approx \StarSED_\ell \, \sum_{k=1}^{K}
  \gamma_\ell^{\beta + k - 1} \, \PSFmode_{k,i} \, .
  \label{eq:PeX-model-with-beta}
\end{equation}
Clearly, if the star SED, $\StarSED(\lambda)$ and the exponent $\beta$ are
both unknown, it is not possible to disentangle them from the resampled
data alone without ambiguities .  We therefore rewrite the model in
Eq.~(\ref{eq:PeX-model-with-beta}) as:
\begin{equation}
  \InterpImg_{i,\ell} \approx \FirstTermSED_\ell \, \sum_{k=1}^{K}
  \gamma_\ell^{k - 1} \, \PSFmode_{k,i} \, ,
  \label{eq:PeX-model2}
\end{equation}
where:
\begin{equation}
  \FirstTermSED_\ell = \StarSED_\ell \, \gamma_\ell^{\beta}
  \label{eq:first-term-sed}
\end{equation}
is the SED of the first significant term of the expansion.

\subsection{Fitting the separable model}
\label{sec:model-fitting}

Assuming independent Gaussian noise for the images in a reference coordinate
system, maximum likelihood estimation of the stellar speckles would be achieved
by minimizing:
\begin{equation}
  \label{eq:multi-mode-cost}
  \chi^2 = \sum_{i,\ell} w_{i,\ell} \,
  \Paren[\Big]{
    \InterpImg_{i,\ell}
    - \FirstTermSED_\ell \sum_{k=1}^{K} \gamma_\ell^{k - 1}
       \, \PSFmode_{k,i}
  }^2 \, ,
\end{equation}
where the statistical weights are given by:
\begin{equation}
  \label{eq:statistical-weights}
  w_{i,\ell} = \begin{cases}
  0 & \text{if $\InterpImg_{i,\ell}$ is unmeasured;}\\
  1/\Var\Brace{\InterpImg_{i,\ell}} & \text{otherwise.}
  \end{cases}
\end{equation}
Taking into account unmeasured data is an important feature as, after alignment
and magnification, the images may have different supports in the considered
reference coordinate system.  To process the noiseless simulated images
considered here, we set the weights to be equal to zero for unseen pixels and
otherwise equal to one.  Because of the resampling of the images, the values of
$\InterpImg_{i,\ell}$ are certainly correlated and this could be taken into
account using non diagonal statistical weights in the expression of the penalty
$\chi^2$.  For the sake of simplicity, we consider independent statistics as assumed by Eq.~(\ref{eq:multi-mode-cost})in
the following.

For a given number of terms $K$ in the expansion, the unknowns of the problem
are the PSF modes, denoted by $\V{\PSFmode}$, and the SED of the first
significant mode, denoted by $\V{\FirstTermSED}$. Fitting our model therefore
amounts to solving the problem:
\begin{equation}
  \label{eq:bilinear-problem}
  \Paren[\big]{\estim{\V{\PSFmode}}, \estim{\V{\FirstTermSED}}} =
  \argmin_{\V{\PSFmode},\V{\FirstTermSED}} \chi^2 \, ,
\end{equation}
where $\chi^2$ is defined in Eq.~(\ref{eq:multi-mode-cost}). Solving this
problem turns out to be a very difficult task because the model is bilinear in
the parameters even though the penalty $\chi^2$ is quadratic with respect to
the model.  Finding one of the components ($\V{\PSFmode}$ or
$\V{\FirstTermSED}$) of the model given the other ($\V{\FirstTermSED}$ or
$\V{\PSFmode}$) is comparatively trivial as it requires to solve a weighted
linear least squares problem.  In practice, it should not be too difficult to
derive an estimation $\estim{\V{\FirstTermSED}}$ of the first mode SED $\V{\FirstTermSED}$ and solving the
difficult problem~(\ref{eq:bilinear-problem}) can be avoided.

\subsection{Approximation by a truncated singular value decomposition}
\label{sec:TSVD}

Clearly, the model in Eq.~(\ref{eq:PeX-model1-compact}) is a separable
approximation of the interpolated distribution.  The singular value
decomposition (SVD) invented by \citet{Eckart_Young-1939-SVD} and
\citet{Mirsky-1960-symmetric_gauge} is the perfect tool to extract a separable
model from the resampled data.  The SVD of $\M{\InterpImg}$ is written:
\begin{equation}
  \InterpImg_{i,\ell} = \sum_{k = 1}^{\mathclap{\Rank(\M{\InterpImg})}}
U_{i,k}\,\sigma_{k}\,V_{\ell,k} \, ,
\end{equation}
or, using matrix notation:
\begin{equation}
  \M{\InterpImg} = \M{U}\cdot\M{\Sigma}\cdot\M{V}\T
  = \sum_{k = 1}^{\mathclap{\Rank(\M{\InterpImg})}}
\sigma_{k}\,\V{u}_{k}\cdot\V{v}_{k}\T
    \, .
\end{equation}
where $\M{U}$ and $\M{V}$ are orthonormal matrices whose $k$-th columns are
$\V{u}_{k}$ and $\V{v}_{k}$, the so-called left and right \emph{singular
vectors} of $\M{\InterpImg}$, and $\M{\Sigma}$ is a diagonal matrix whose
diagonal elements are called the \emph{singular values} of $\M{\InterpImg}$
denoted by $\sigma_k$.  By convention, the singular values are all nonnegative
and sorted in descending order:
\begin{displaymath}
  \sigma_1 \ge \sigma_2 \ge \ldots \ge \sigma_{\Rank(\M{\InterpImg})} > 0 \, ,
\end{displaymath}
and all singular values for $k > \Rank(\M{\InterpImg})$ are equal to zero.

According to the Eckart-Young-Mirsky theorem
\citep{Eckart_Young-1936-approximation, Mirsky-1960-symmetric_gauge}, the SVD
truncated to the first $K$ singular modes, provides the best approximation of
this rank to the original matrix $\M{\InterpImg}$ in a least squares sense.
Thus no other bilinear (separable) model with $K$ modes can beat the one built
from the truncated SVD. Approximating the resampled data by the truncated
singular value decomposition (TSVD) is written:
\begin{equation}
  \label{eq:tsvd}
  \M{\InterpImg} \approx
  \sum_{k = 1}^{K} \sigma_k\,\V{u}_{k}\cdot\V{v}_{k}\T \, ,
\end{equation}
with $K \le \Rank(\M{\InterpImg})$.

Using the SVD to determine a separable model is not new, it is for instance the
method of choice to perform the principal component analysis (PCA) of data. The
SVD has however some limitations in our context: (i) it yields the optimal
separable decomposition in an ordinary least squares sense but cannot cope with
statistical weights or missing data\footnote{even though it is possible to
discard some \emph{bad pixels} but this has to be done for all spectral
channels at the same interpolated locations, that is by removing some rows of
the data matrix $\M{\InterpImg}$}; (ii) it requires working with the
interpolated data $\M{\InterpImg}$; (iii) it does not include any \emph{a
priori} behavior that can be dictated by the physics and which could be
introduced to improve the estimation.  The interpolated data are necessarily
correlated while measurements in the raw data may be statistically independent.
Using ordinary least squares is suboptimal compared to weighted least squares
which can also cope with missing data. Finally, the series expansion based on
physical considerations (diffraction) shows that the chromatic weights are
fairly well constrained, while such constraints cannot be imposed in an
SVD-based analysis.

SVD can however be used to investigate the chromatic behavior of the
distribution of speckles and to provide guidelines to design a more restrictive
separable model as well as initial parameters for this model. This model can
then be used to fit the data in an inverse approach in order to relieve all the
drawbacks of SVD.  Besides, since TSVD directly yields the best approximation
of this rank, it can serve as a template to evaluate the precision achieved by
any other approximation such as that in Eq.~(\ref{eq:PeX-model1-compact}).  Due
to the coronographic mask in the image plane, the model derived from the series
expansion cannot apply everywhere and is certainly wrong in the central region
--- this is mostly critical for SVD.  This is why, unless explicitly stated
otherwise, we exclude the central region of the field of view in our subsequent
SVD based analysis.

\subsection{Data simulation}
\label{sec:simulation}

In order to check the proposed separable approximation of the coronographic
images, we simulated multi-spectral images with parameters typical of
\noun{Sphere} \citep{Beuzit_et_al-2010}: $8.2\,\meter$ telescope, equipped with
a Lyot coronograph with an apodized pupil as described by
\citet{Carbillet_et_al-2011-apodized_coronograph} and a pixel size of
$7.4\,\text{milliarcseconds}$ (mas).  Pixel integration was taken into account
assuming a $100\%$ fill factor.  We considered $21$ spectral channels evenly
distributed over the H band ($1.46-1.84\,\micro\meter$).  To account for
imperfect wavefront correction, we introduced rather pessimistic phase
aberrations of $70\,\nano\meter$ rms with the same powerspectrum as measured on
\noun{Sphere}.  Our simulations did not include filtering of turbulence-induced
phase errors by the AO system.


Typical simulated images are shown in Fig.~\ref{fig:images}. The level of
the brightest speckles in these images shows that a contrast of greater
than $10^{-3} $ is needed to detect a planet using one of these
coronographic images.  Given these simulated images, we resampled the
images in the different spectral channels to compensate for the chromatic
magnification (see rightmost image of Fig.~\ref{fig:images}).  Testing the
model on the resampled data cube is described in the following sections.

\subsection{Behavior of the most significant mode}
\label{sec:first-mode}

\begin{figure}
  \centering \includegraphics[scale=0.60]{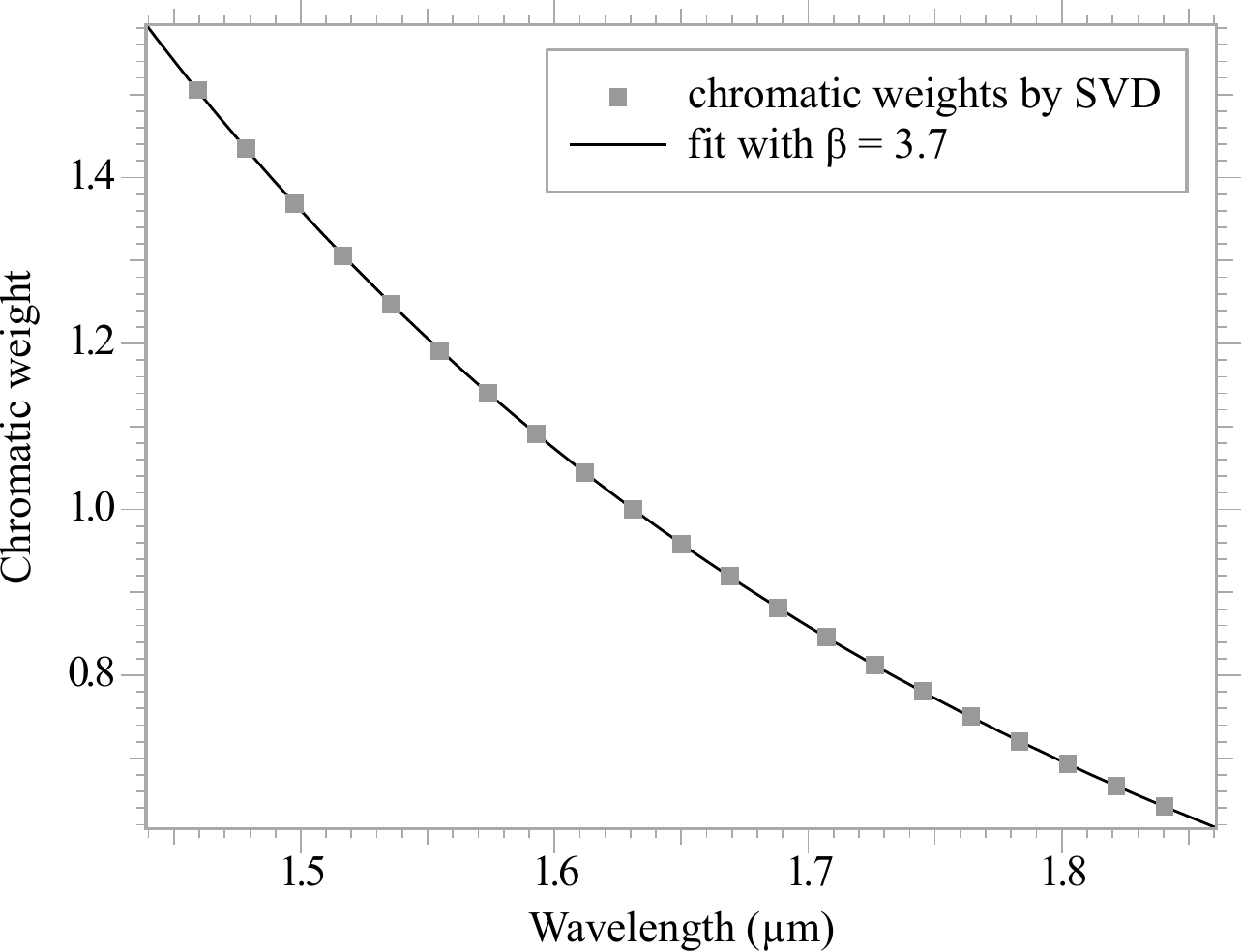}
  \caption{Chromatic weights computed by SVD.  The symbols show the chromatic
  weights for the first mode of the SVD decomposition of the 21 spectral
  channels of the \noun{Sphere}-like simulation for an angular distance
  $\theta\ge0.25$\,arcsec. The curve is the result of a least square fit of a
  power law $\eta\,(\lambda^\Tag{ref}\!/\lambda)^\beta$ with $\beta \simeq
  3.7$.} \label{fig:chromatic-weights}
\end{figure}

Comparing the TSVD factorization in Eq.~(\ref{eq:tsvd}) with our model in
Eq.~(\ref{eq:PeX-model1-compact}) yields the following correspondences:
\begin{subequations}
\begin{align}
  \V{\PSFmode}_{k} &\approx \alpha_k \, \V{u}_{k} \, , \\
  \V{\SpctWgt}_{k} &\approx (\sigma_k/\alpha_k) \, \V{v}_{k} \, ,
\end{align}
\end{subequations}
for some arbitrary factors $\alpha_k \not= 0$ which must be introduced because
the SVD modes $\V{u}_k$ are not normalized in the same way as our PSF modes
$\V{\PSFmode}_{k}$.  Our model imposes more constraints than SVD does, and the
above relations are therefore unlikely to be matched exactly.  It is however
interesting to investigate whether TSVD and our model yield similar results in
the case of a single mode approximation\footnote{This would be a clear
indication that the physical constraints correctly capture the relevant
information.}.  In this case, we expect that:
\begin{displaymath}
   \Paren{\V{v}_{1}}_{\ell}/\StarSED_\ell
   \approx (\alpha_1/\sigma_1) \,
           \Paren{\V{\SpctWgt}_{1}}_{\ell}/\StarSED_\ell
   \propto \gamma_{\ell}^{\beta}
   \, ,
\end{displaymath}
with $\StarSED_\ell = \StarSED(\lambda_\ell)$, $\gamma_\ell =
\lambda^\Tag{ref}\!/\lambda_\ell$ and where $\beta = k_0 + 2$ is the chromatic
exponent associated with the first significant mode.
Figure~\ref{fig:chromatic-weights} shows the values of
$\Paren{\V{v}_{1}}_{\ell}/\StarSED_\ell$ found by the SVD decomposition of the
speckles of our simulation interpolated in a reference frame and for angular
distance $\theta\ge0.25$\,arcsec.  Clearly, a power law (the curve in
Fig.~\ref{fig:chromatic-weights}):
\begin{equation}
   \Paren{\V{v}_{1}}_{\ell}/\StarSED_\ell \approx
   \eta \, \gamma_{\ell}^{\beta} \, ,
   \label{eq:power-law}
\end{equation}
with $\beta \simeq 3.7$ provides a perfect fit of the spectral weights
estimated by SVD.  This agreement between the spectral weights found by SVD
(which makes no specific assumptions about their chromatic behavior) and the
power law induced from the diffraction is a first validation of the proposed
chromatic model.

\begin{figure}
  \centering \includegraphics[scale=0.4]{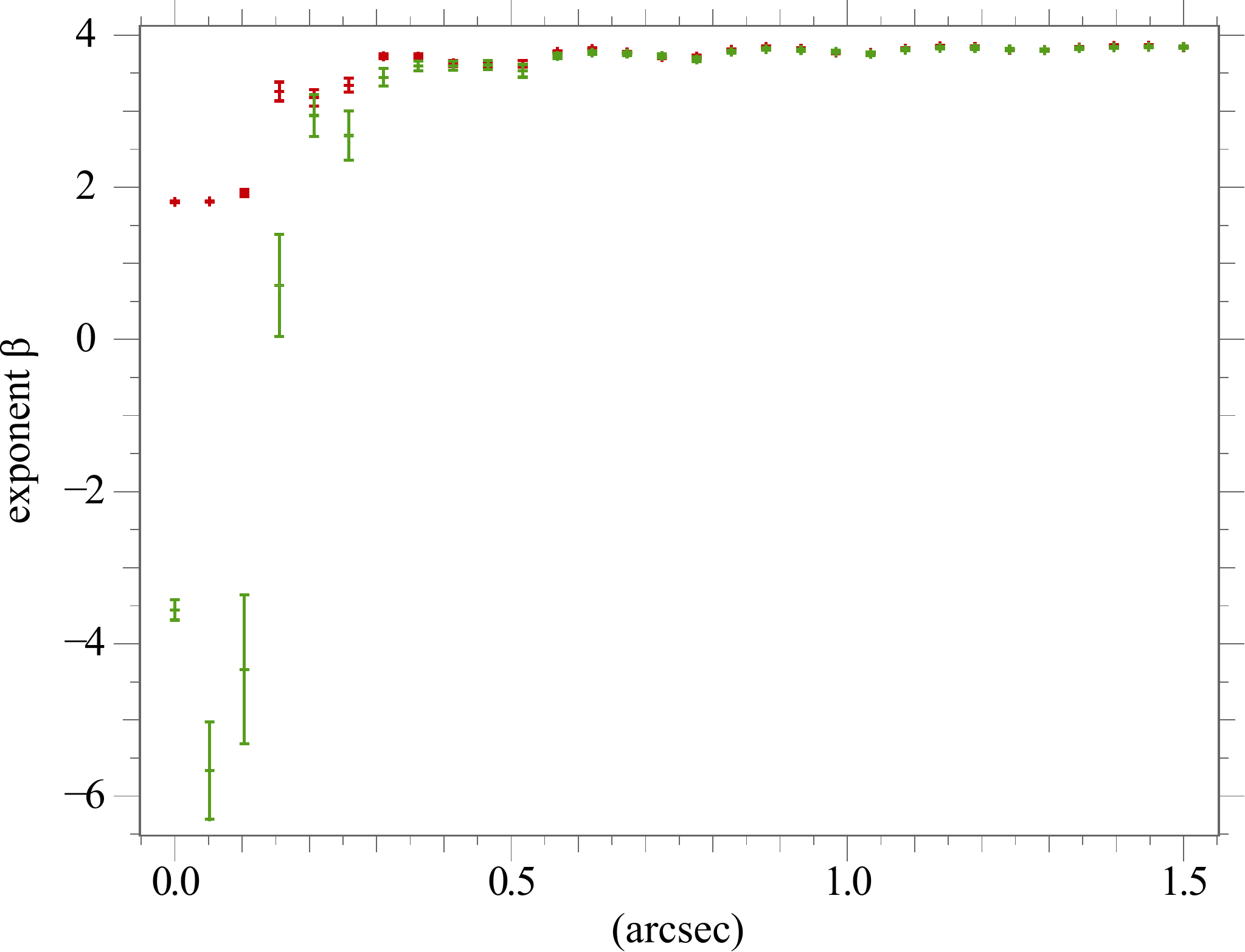}

  \caption{Chromatic exponent $\beta$.  The spectral weights computed by SVD on
  annular regions (of thickness $0.05$\,arcsec) of the rescaled hyperspectral
  cube have been fitted by a power law $(\lambda^\Tag{ref}\!/\lambda)^\beta$.
  The points show the average values and their error bars computed from 50
  \noun{Sphere}-like simulations in the H band with a 70\,nm rms residual
  aberration.  The red symbols are for a simulation with no coronograph, the
  green symbols are with an apodized Lyot coronograph.}
  \label{fig:chromatic-exponent}
\end{figure}

In order to investigate whether the spectral behavior depends on the angular
distance, we extracted narrow annular regions (centered on the star position)
from the interpolated cubes and computed the SVD of these data subsets.  The
chromatic behavior of the first SVD right singular mode, $\V{v}_1$, of these
decompositions is then fit with the power law in Eq.~(\ref{eq:power-law}). The
exponents $\beta$ obtained for different angular distances $\theta$ from the
center are plotted in Fig.~\ref{fig:chromatic-exponent} (two cases are
considered: with and without a coronograph).  Without a coronograph, the
exponent is $\beta\simeq2$ for $\theta \le 0.1\arcsec$ which is exactly what is
expected from diffraction in the aberration-free regime.  For larger distances,
the exponent grows rapidly to a flat level $\beta \simeq 3.7$ due to the
aberrations\footnote{We checked that without a coronograph or aberrations, the
exponent is $\beta\simeq2$ everywhere.}.  With a coronograph, the exponent is
very different near the center where it can be as small as $\beta\simeq-6$
depending on the realization of the random aberrations; around the distance
$\theta\simeq0.15\arcsec$, the exponent grows rapidly to reach the same plateau
at $\beta \simeq 3.7 \pm 0.1$ as in the case with no coronograph.

The curves in Fig.~\ref{fig:chromatic-weights} and \ref{fig:chromatic-exponent}
indicate that the best exponent for the $\theta \ge 0.2 \arcsec$ region is thus
$\beta^\star \simeq 3.7 \pm 0.1$ and we observed the same behavior with various
aberration levels (in the range $60-100\,\nano\meter$).  Remembering that for
the first SVD mode, we should have $\beta^\star = k_0 + 2$ with $k_0$ the index
of the most significant term in the chromatic expansion (\ref{eq:interp-image})
and noting that $\beta^\star \simeq 3.7$ is close to $4$ but is not integer, we
deduce that $k_0 = 2$ is the most significant mode in the model given in
Eq.~(\ref{eq:interp-image}) but that other modes are needed to correctly
approximate the actual speckle pattern.  The most significant mode has an even
order, and indeed the most prominent speckles seems to be symmetrically
distributed in Fig.~\ref{fig:images}.  The index, $k_0=2$, of the first
significant mode is an indication of the efficiency of the coronograph.

To support this deduction, we compared the results of the single mode TSVD
approximation which does not implement any specific chromatic behavior with
our model given in Eq.~(\ref{eq:PeX-model-with-beta}) with $K=1$.  With a
single mode and assuming the star SED and the chromatic exponents are known, the
solution to minimizing $\chi^2$ defined in Eq.~(\ref{eq:multi-mode-cost}) with
respect to $\V{\PSFmode}_1$ is given trivially by:
\begin{equation}
  \estim{\PSFmode}_{1,i} = \frac{
    \sum_{\ell}  w_{i,\ell} \, \StarSED_\ell \, \gamma_\ell^\beta \,
    \InterpImg_{i,\ell}
  }{
    \sum_{\ell}  w_{i,\ell} \,  \StarSED_\ell^2 \, \gamma_\ell^{2\,\beta}
  } \, .
  \label{eq:pex1-solution}
\end{equation}

\begin{figure*}
  \centering \includegraphics[scale=0.45]{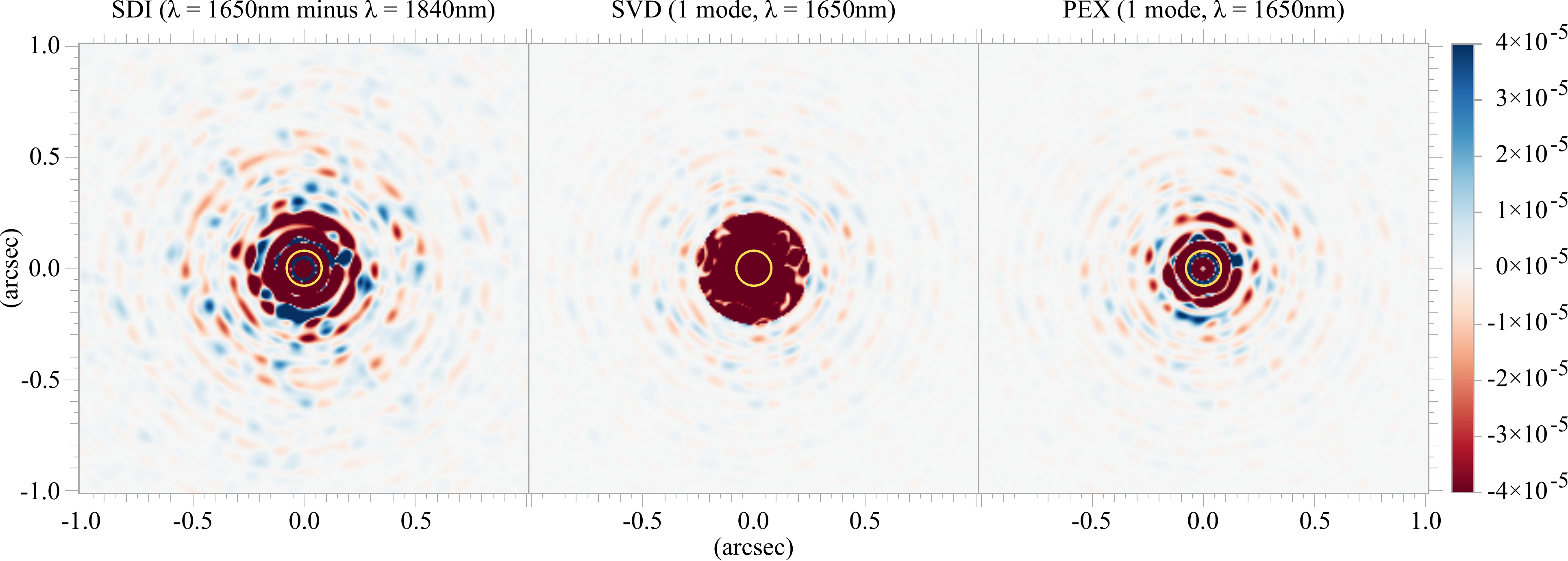}

  \caption{Residuals for a single mode separable approximation of the
  multi-spectral data.  The residuals are shown for the $\lambda =
  1650\,\nano\meter$ channel and have been normalized by the peak intensity
  without a coronograph. \emph{Left}: residuals by SDI; \emph{center}:
  residuals with truncated SVD; \emph{right}: residuals with spectral
  weighting set to $(\lambda^\Tag{ref}\!/\lambda)^{3.7}$ and a linear least
  square fit of the spatial mode.  The coordinates are in arc-seconds
  (interpolated at the shortest wavelength).  The values have been clipped
  to the range $[-4\times10^{-5},+4\times10^{-5}]$ and color scales are the
  same for all sub-figures: blue for positive residuals, red for negative
  residuals and levels in relative contrast units.  The coronograph mask is
  indicated by a yellow circle.  See the text for details of the
  simulation.} \label{fig:single-mode-separable-model}
\end{figure*}

\begin{figure}
  \centering
  \includegraphics[scale=0.6]{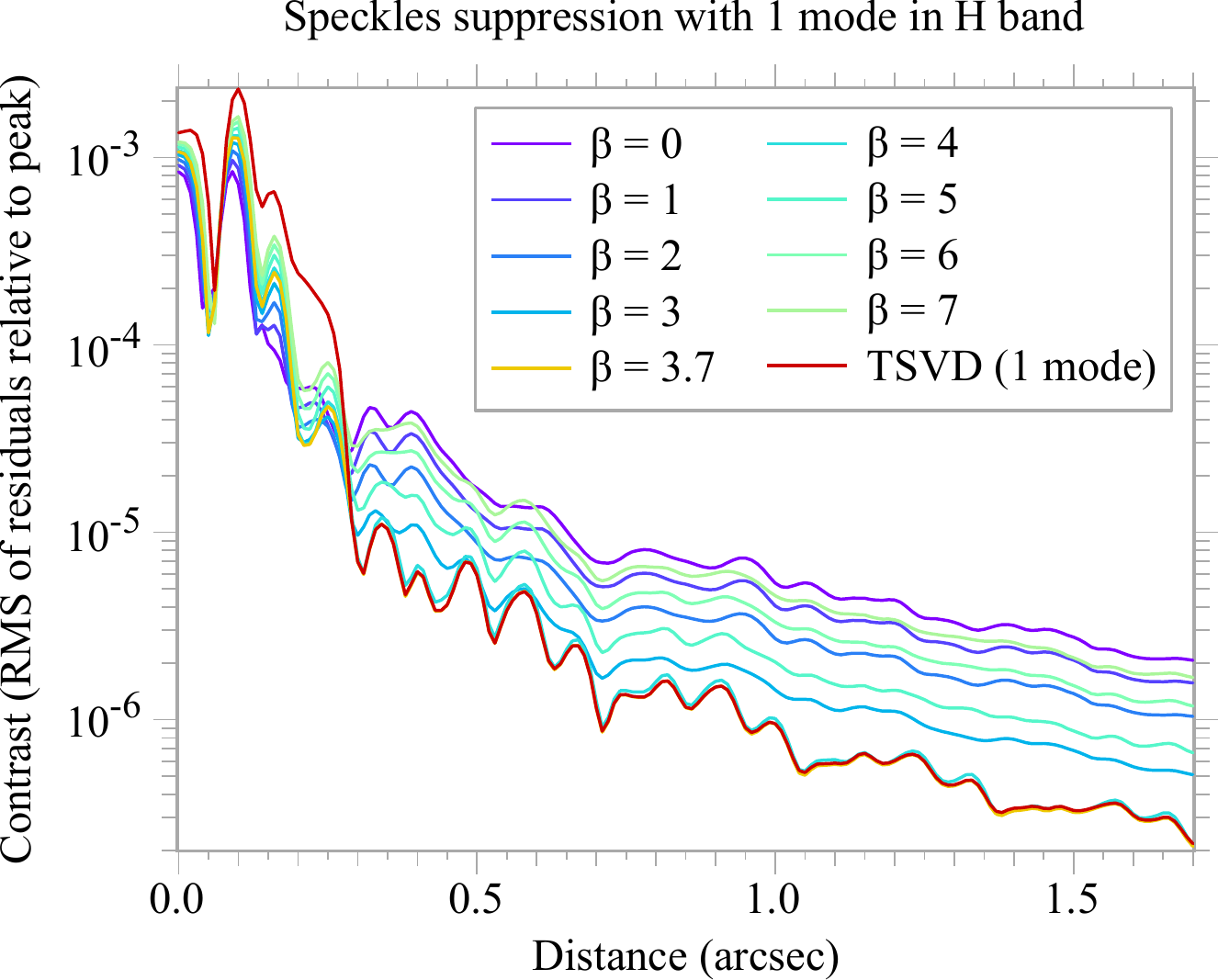}

  \caption{Residuals for a single mode approximation of the multi-spectral
  distribution by truncated SVD (TSVD) or by the proposed model:
  $(\lambda^\Tag{ref}\!/\lambda)^\beta \,\PSFmode(\RefPos)$ and various
  values of the exponent $\beta$.  The computations were done for
  conditions typical of that of \noun{Sphere} in the H band, see the text
  for details.} \label{fig:rms-profile-pex1}
\end{figure}

\newcommand{\ProfilePlot}[2]{\begin{figure}
  \centering \includegraphics[scale=0.6]{#2}

  \caption{Residuals for a #1-mode approximation of the multi-spectral
  distribution by truncated SVD (TSVD) or by the proposed model:
  $\sum_{k=1}^{#1}(\lambda^\Tag{ref}\!/\lambda)^{\beta_k}
  \,\PSFmode_k(\RefPos)$.  The conditions are identical to those of
  Fig.~\ref{fig:rms-profile-pex1}.} \label{fig:rms-profile-pex#1}
\end{figure}}

Figures~\ref{fig:single-mode-separable-model} and \ref{fig:rms-profile-pex1}
show the efficiency of the speckle suppression by different single mode
approximations: SDI, TSVD (with a mask to discard the central $\theta <
0.25\,\arcsec$ region whose chromatic behavior is very different from the other
parts of the field of view) and our model computed according to
Eq.~(\ref{eq:pex1-solution}).  In Fig.~\ref{fig:single-mode-separable-model} we
took $\beta = 3.7$ while different values of $\beta$ are considered in
Fig.~\ref{fig:rms-profile-pex1}. In order to express the results in terms of
the planet/star contrast achieved, the residuals have been normalized by the
peak intensity in the image computed under the same conditions but with no
coronograph.  The profiles plotted in Fig.~\ref{fig:rms-profile-pex1} are the
root mean squared (rms) value of the normalized residuals for a given angular
distance for all spectral channels (\ie averaging is carried out for all
azimuthal angles and wavelengths).  At this point, only the distribution of
speckles is considered, there is no noise in the simulations.  The curves
presented in Fig.~\ref{fig:rms-profile-pex1} therefore really measure the
ability of the various approximations to remove the stellar leakage.
Figure~\ref{fig:rms-profile-pex1} shows that the best speckle suppression is
achieved when $\beta\rightarrow\beta^\star\simeq3.7$ as predicted from
Fig.~\ref{fig:chromatic-exponent}.  Taking $\beta=4$ is nearly as good as with
$\beta = \beta^\star$ but any other integer values for $\beta$ yield
significantly worse results. Finally, the level of contrast achieved assuming a
specific chromatic behavior (with a suitable exponent $\beta^\star$) is as good
as what is obtained by TSVD which validates our approach.
Figure~\ref{fig:single-mode-separable-model} clearly shows that the single mode
TSVD and \PeX methods yield similar residuals in both distribution and magnitude
for angular distances $\theta \ge 0.25\,\arcsec$ where they are better than
SDI.  In our model, interpolated pixels are fitted independently, thus,
contrary to TSVD, no masking of the central region is needed by \PeX. Even
though the residuals remain important in the central region, \PeX is able to
reduce the speckles for smaller angular distances than SVD.

As already noted, the speckles in the unprocessed images
(Fig.~\ref{fig:images}) look mostly symmetrical in agreement with the even
order of the most significant mode.  On the contrary, the residual images in
Fig.~\ref{fig:single-mode-separable-model} appear to be nearly antisymmetrical.
 This is consistent with the diffraction based expansion: as an even mode ($k_0
= 2$) has been removed, the next most significant mode should be an odd order
mode at $k_0\pm1$ and thus antisymmetrical.  We believe that this further
supports the model in Eq.~(\ref{eq:PeX-model2}) and we examine the performance
when more modes are removed.

\subsection{Fitting multiple modes}

In order to improve the speckle suppression, we now consider fitting
more than one mode.  As shown by the previous section, applying TSVD (with a
single mode) to the resampled images yields a good estimate of the chromatic
weights of the first PSF mode.  Indeed Eq.~(\ref{eq:power-law}) yields:
\begin{equation}
  \FirstTermSED_\ell = \StarSED_\ell \, \gamma_\ell^{\beta}
  \approx \eta\,\Paren{\V{v}_{1}}_{\ell} \, ,
  \label{eq:first-mode-sed-approx}
\end{equation}
where $\V{v}_{1}$ is the first right singular vector of the SVD decomposition
of the resampled data and $\eta$ is an irrelevant normalization factor.  Then,
as noted before, the remaining unknowns of the problem, namely the PSF modes
$\V{\PSFmode}$, are easily found as the result of a weighted linear least
squares fit:
\begin{equation}
  \estim{\V{\PSFmode}} = \argmin_{\V{\PSFmode}}
  \sum_{i,\ell} w_{i,\ell} \,
  \Paren[\Big]{
    \InterpImg_{i,\ell}
    - \FirstTermSED_\ell \sum_{k=1}^{K} \gamma_\ell^{k - 1}
       \, \PSFmode_{k,i}
  }^2 \, .
  \label{eq:best-psf-modes}
\end{equation}
Compared to the problem in Eq.~(\ref{eq:multi-mode-cost}), the new problem is
trivial to solve.  An additional simplification is that the problem is
separable with respect to the spatio-temporal index $i$.  In other words, for
each spatio-temporal sample $(\RefPos_i,t'_i)$, one has to solve:
\begin{equation}
  \Brace{\estim{p}_{k,i}}_{k=1,\ldots,K} = \argmin_{\V{x}\in\Reals^K}
  \sum_{\ell} w_{i,\ell} \,
  \Paren[\Big]{
    \InterpImg_{i,\ell}
    - \FirstTermSED_\ell \sum_{k=1}^{K} \gamma_\ell^{k - 1} \, x_{k}
  }^2 \, ,
  \label{eq:separable-psf-fit}
\end{equation}
which amounts to solving a linear system of only $K$ unknowns (for each
index $i$).  This is similar to the \emph{spectral deconvolution} method
proposed by \citet{Sparks_Ford-2002-imaging_spectroscopy} except that the
chromatic exponents are not the same and that all terms are multiplied by a
common SED $\V{\FirstTermSED}$.

To compare the multi-mode \PeX model with TSVD, we assume again that the
stellar SED is known and consider spectral exponents $\beta_k = k_0 + k - 1$
for different values of the index $k_0$ of the most significant mode.  The
estimated PSF modes are then given by:
\begin{displaymath}
  \Brace{\estim{p}_{k,i}}_{k=1,\ldots,K} = \argmin_{\V{x}\in\Reals^K}
  \sum_{\ell} w_{i,\ell} \,
  \Paren[\Big]{
    \InterpImg_{i,\ell}
    - \StarSED_\ell \sum_{k=1}^{K} \gamma_\ell^{\beta_k} \, x_{k}
  }^2 \, .
\end{displaymath}
Figure~\ref{fig:rms-profile-pex2} shows the rms level of the residuals after
subtracting the 2-mode models from the simulated images. Compared to TSVD, the
\PeX model achieves the same efficiency providing the correct spectral exponents
are selected.  In this case, $\V{\beta} = \{3,4\}$ or $\V{\beta} = \{4,5\}$ are
the best  and correspond respectively to $k_0 = 3$ or $4$.  Compared to the
unprocessed coronographic images, the gain is $\sim 10^{-7}$ in contrast;
compared to the single mode models, there is a $\sim10$ factor improvement.
Figure~\ref{fig:residuals-with-2-modes} displays the residuals in the $\lambda
= 1650\,\nano\meter$ spectral channel.  For an angular distance $\theta \ge
0.25\arcsec$ the residuals by TSVD and by the proposed method have almost
exactly the same distribution.  The central part has been masked for TSVD which
therefore performs poorly compared to \PeX.

With 3 modes, Fig.~\ref{fig:rms-profile-pex3} shows that the efficiency of \PeX
no longer depends on the specific choice of the spectral exponents (even
though we only checked for a limited range: $2 \le k_0 \le 6$).  Except
in the central part of the field of view, TSVD performs slightly better than
\PeX (both with 3 modes). This can be seen in the images of the residuals
shown by Fig.~\ref{fig:residuals-with-3-modes}.  Compared to the 2-mode
models, the supplementary mode gains a $\sim3-5$ factor in the reduction of the
level of the residuals depending on the distance from the center.

\begin{figure}
  \centering \includegraphics[scale=0.6]{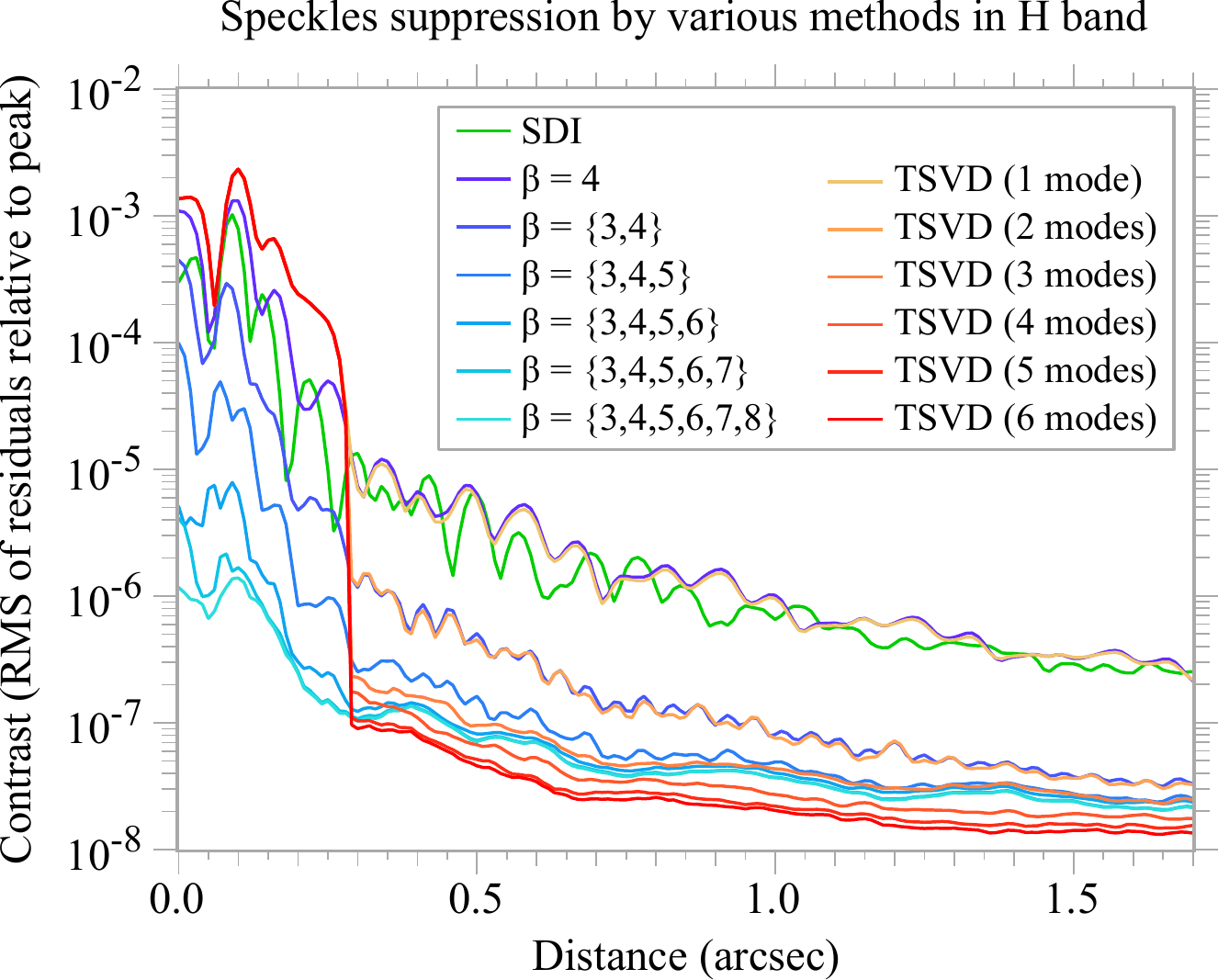}
  \caption{Residuals for different approximations of the multi-spectral
  distribution.  SDI stands for simple resampled image subtraction; the curves
  labeled with $\beta = \ldots$ are for our model with 1 to 6 modes (the
  values are the chromatic exponents); TSVD is for truncated SVD. See
  Fig.~\ref{fig:rms-profile-pex1}--\ref{fig:rms-profile-pex3} and text for
  more details.} \label{fig:rms-profiles}
\end{figure}

Figure~\ref{fig:rms-profiles} summarizes the performances of SDI, TSVD and the
proposed spectral expansion, where the two latter methods are used with different
numbers of modes (from 1 to 6).  With more than 3 modes, we assumed that, as in
the 3-mode case, the efficiency does not so much depend on the list of exponents
and we take $k_0=3$ as the index of the first significant mode which
corresponds to chromatic exponents $\V{\beta} = \Brace{3,4,5,\ldots}$. With a
given number of modes (one for SDI) the different methods have roughly the same
performance, increasing the number of modes improves significantly the
achieved contrast: using 2 modes instead of a single mode improves the detection
contrast by more than an order of magnitude.  Increasing the number of modes
also seems to flatten the level of the residuals as a function of the distance.
With 6 modes, a contrast of $\sim2\times10^{-8}$ is reached at distances larger
than $1\arcsec$ from the center. Performances are worse near the center but a
contrast as low as $10^{-7}$ seems to be reachable near the edges of the mask
with our method. This is very important for the detection of close companions.

\newcommand{\ResidualsPlot}[3]{ %
\begin{figure}
  \centering
  \includegraphics[scale=0.45]{#3}

  \caption{Residuals for a separable approximation of the multi-spectral data
  with #1 modes.  This figure is similar to
  Fig.~\ref{fig:single-mode-separable-model}, except that #1 modes have been
  used with respective spectral weighting set to
  $(\lambda^\Tag{ref}\!/\lambda)^{\beta_k}$ with $\V{\beta} = \{#2\}$ for
  the linear least square fit in the bottom panel.}
  \label{fig:residuals-with-#1-modes}
\end{figure}
}

\ProfilePlot{2}{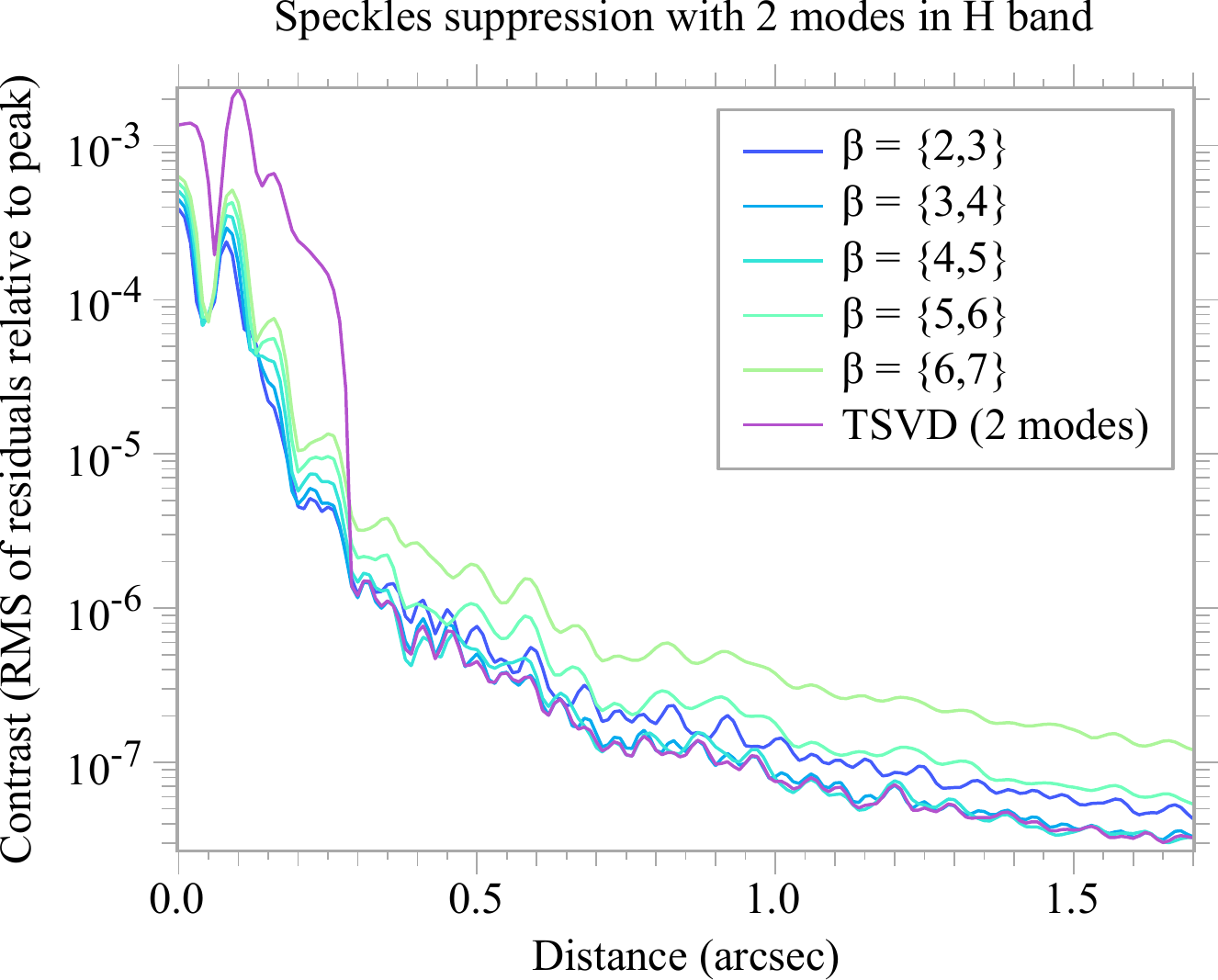}

\ResidualsPlot{2}{3,4}{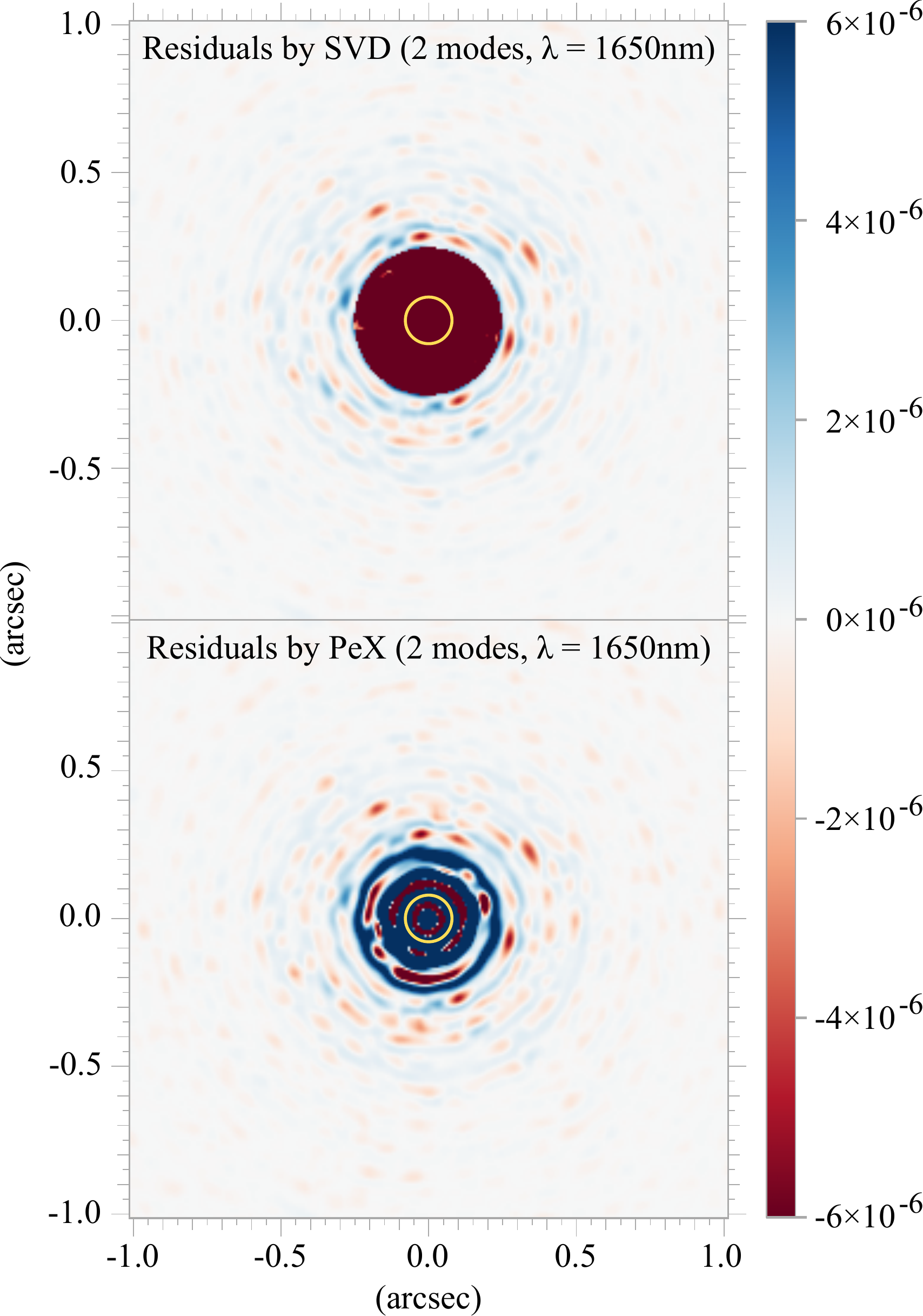}

\ProfilePlot{3}{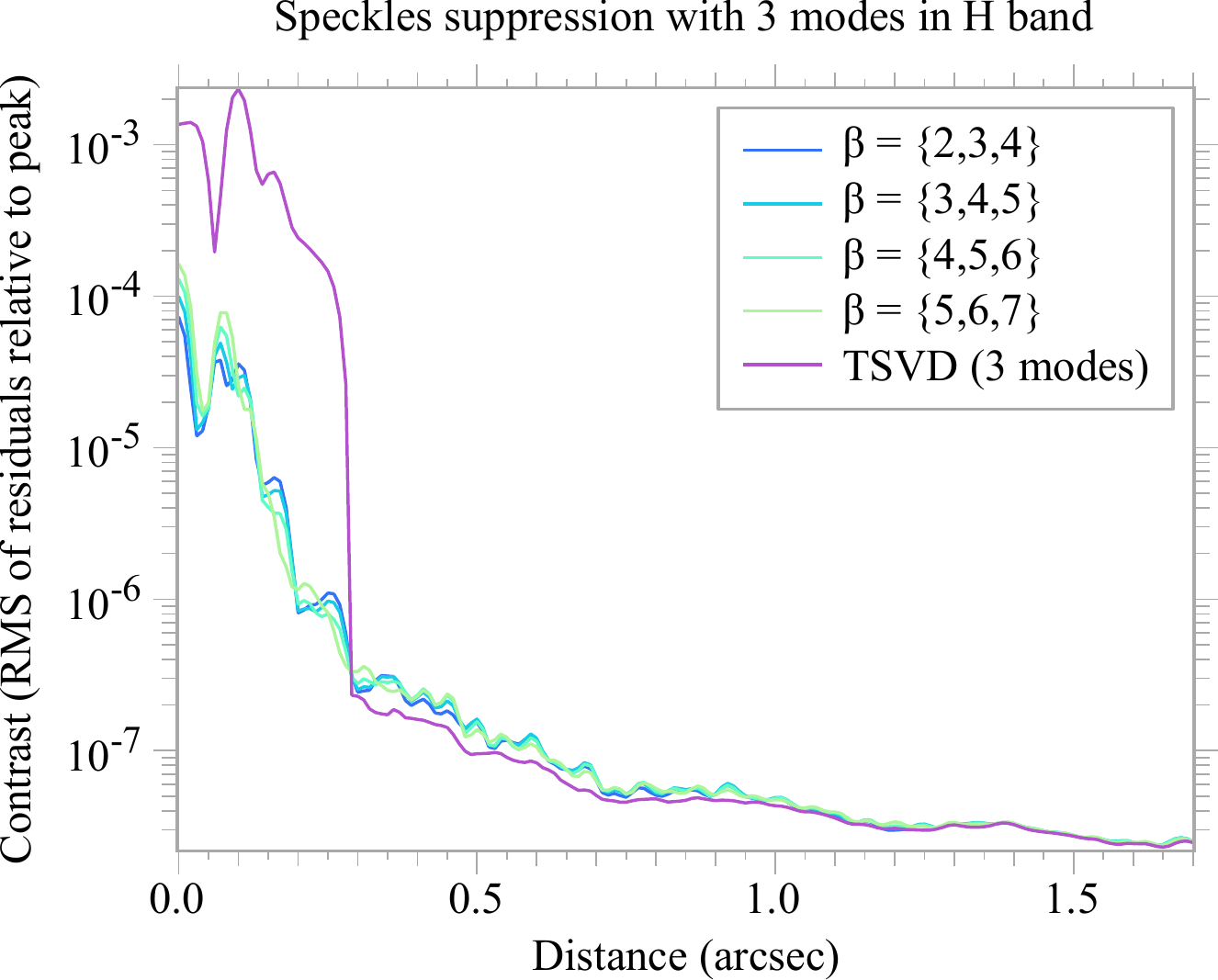}

\ResidualsPlot{3}{3,4,5}{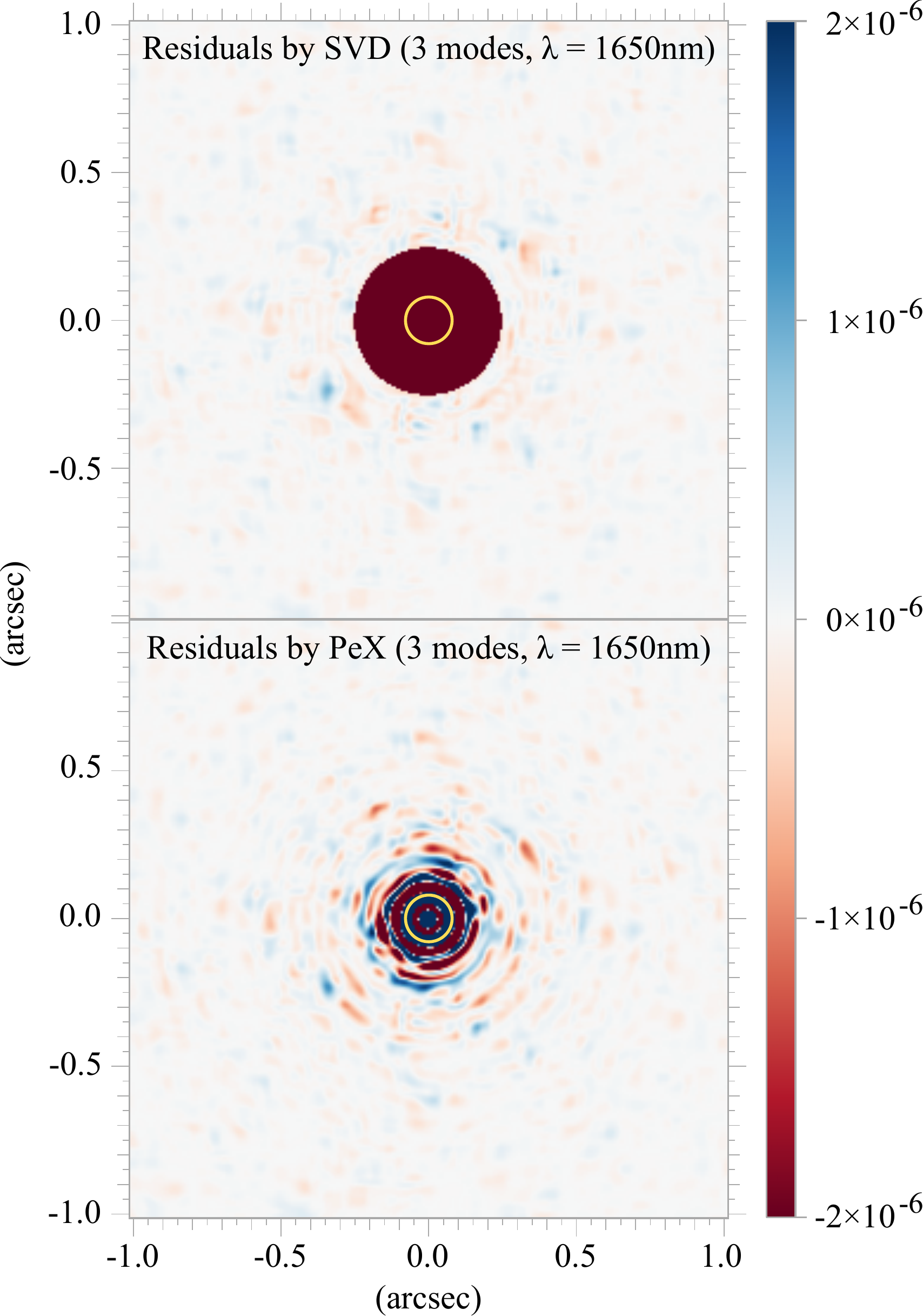}

\section{Application to exoplanet detection}
\label{sec:planet-detection}

Assuming that the planet brightness is negligible compared to that of the
stellar speckles and because the planet position does not vary with wavelength,
our method for speckle removal should be rather insensitive to the presence of
very faint planets.  Even though a better approach that we will consider in a
following paper would be to \emph{jointly} perform speckle removal and planet
detection, it is tempting to perform planet detection in the residuals obtained
by subtracting the fitted model of the speckles from the observed images.  This
task is considered in this section.  We first derive a detection test that can
be applied to multi-variate data (the considered images depend on the
wavelength and on the exposure); we then apply this test to simulated data.

\subsection{Criterion for detection in multi-variate data}
\label{sec:detection-criterion}

Planet detection amounts to deciding between two hypotheses: no planet is
present ($\mathscr{H}_0$) or a planet is present ($\mathscr{H}_1$).  If there
is a planet at position $\ImgPos$ (hypothesis $\mathscr{H}_1$), the model of
the image after removal of speckles is:
\begin{equation}
  y_{j,\ell,m} = \PlanetSED_{\ell}
   \, h(\ImgPos_{j,\ell,m} - \ImgPos, \lambda_\ell)
   + \varepsilon_{j,\ell,m} \, ,
   \label{eq:residuals-model}
\end{equation}
where $y_{j,\ell,m}$ is the value of the residual image at the $j$-th pixel of
the $\ell$-th spectral channel and the $m$-th exposure, $\PlanetSED_\ell$ is
the planet flux at wavelength $\lambda_\ell$ of the considered spectral
channel, $h(\ImgPos, \lambda)$ is the off-axis PSF at sky position $\ImgPos$
and wavelength $\lambda$, and the term $\varepsilon_{j,\ell,m}$ accounts for
the noise (and model errors).  Of course, if there are no planets, then
$\PlanetSED_{\ell} = 0$ ($\forall\ell$) and the residuals are just due to the
noise and the model under hypothesis $\mathscr{H}_0$ is just:
\begin{equation}
  y_{j,\ell,m} = \varepsilon_{j,\ell,m} \, .
\end{equation}
Deciding between the two hypotheses can be based on the level of the
generalized likelihood ratio \citep[GLR, see \eg][]{Kay-1998-detection_theory}
which is:
\begin{equation}
   \mathrm{GLR}(\ImgPos) = \frac{
      \max_{\V{\PlanetSED}}\Pr(\Vy \given \ImgPos, \V{\PlanetSED}, \mathscr{H}_1)
   }{
      \Pr(\Vy \given \mathscr{H}_0)
   } \, ,
   \label{eq:GLR}
\end{equation}
where $\Pr(\Vy \given \ldots)$ is the likelihood of the residuals $\Vy$
conditioned by the knowledge of some information or parameters represented by
the ellipsis.  The higher the GLR, the more likely is a detection and
vice-versa.  For some chosen threshold $\tau$, this is summarized by the
notation:
\begin{displaymath}
   \mathrm{GLR}(\ImgPos)
   \mathop{\gtrless}_{\mathscr{H}_0}^{\mathscr{H}_1} \tau \, ,
   \label{eq:GLR-test}
\end{displaymath}
which means that $\mathscr{H}_1$ is decided if $\mathrm{GLR}(\ImgPos) > \tau$,
while $\mathscr{H}_0$ is decided if $\mathrm{GLR}(\ImgPos) < \tau$.

For Gaussian independent noise, the cologarithm of the GLR is:
\begin{align*}
   \Lambda(\ImgPos)
   &= -\log\mathrm{GLR}(\ImgPos) \notag \\
   &= \frac{1}{2} \, \Bigl\{
     \sum_{j,\ell,m} w_{j,\ell,m} \, y_{j,\ell,m}^2 \notag \\
   &\hspace*{3em}
   - \min_{\V{\PlanetSED}}
   \sum_{j,\ell,m} w_{j,\ell,m} \, \Paren{
     y_{j,\ell,m} - \PlanetSED_{\ell} \,
     h_{j,\ell,m}(\ImgPos)
   }^2 \Bigr\} \, ,
\end{align*}
where:
\begin{displaymath}
  h_{j,\ell,m}(\ImgPos) = h(\ImgPos_{j,\ell,m} - \ImgPos, \lambda_\ell) \, ,
\end{displaymath}
and $w_{j,\ell,m} \ge 0$ are statistical weights.  Following the reasoning
leading to Eq.~(\ref{eq:statistical-weights}), the weights are given by:
\begin{equation}
  w_{j,\ell,m} = \begin{cases}
  0 & \text{for unmeasured values,} \\
  1/\Var\Brace{y_{j,\ell,m}} & \text{otherwise.}
  \end{cases}
  \label{eq:statistical-weights}
\end{equation}
Expanding and simplifying $\Lambda(\ImgPos)$ yields:
\begin{equation}
  \Lambda(\ImgPos)
  = \sum_{\ell} \min_{\PlanetSED_\ell} \Brace*{
      b_\ell(\ImgPos) \, \PlanetSED_\ell
      - \frac{1}{2} \, a_\ell(\ImgPos) \, \PlanetSED_\ell^2
    }  \, ,
  \label{eq:colog-GLR-1}
\end{equation}
with:
\begin{subequations}
\begin{align}
   a_\ell(\ImgPos)
   &= \sum_{j,m} w_{j,\ell,m} \, h_{j,\ell,m}^2(\ImgPos) \, ,
   \label{eq:glr-den}   \\
   b_\ell(\ImgPos)
   &= \sum_{j,m} w_{j,\ell,m} \, y_{j,\ell,m} \, h_{j,\ell,m}(\ImgPos) \, .
   \label{eq:glr-num}
\end{align}
\end{subequations}
The above expression for $\Lambda(\ImgPos)$ shows that obtaining the maximum
likelihood estimator (MLE) of the planet SED (assuming the planet position) is
simply a matter of solving separable simple quadratic problems for each
spectral channel and yields:
\begin{align}
   \estim{\PlanetSED}_\ell(\ImgPos)
   &= \argmin_{\PlanetSED_\ell} \Brace*{
      b_\ell(\ImgPos) \, \PlanetSED_\ell
      - \frac{1}{2} \, a_\ell(\ImgPos) \, \PlanetSED_\ell^2
    } \notag \\
   &= \frac{b_\ell(\ImgPos)}{a_\ell(\ImgPos)} \, .
   \label{eq:MLE-planet-SED}
\end{align}
Substituting this result in Eq~(\ref{eq:colog-GLR-1}) gives:
\begin{equation}
  \Lambda(\ImgPos) =
  \frac{1}{2} \, \sum_{\ell} \frac{b_\ell^2(\ImgPos)}{a_\ell(\ImgPos)} \, .
  \label{eq:colog-GLR-2}
\end{equation}
The term $a_\ell(\ImgPos)$ does not depend on the data, while the term
$b_\ell(\ImgPos)$ does depend on the data.  The variance of this latter term
can be computed as follows:
\begin{align*}
  \Var\Brace{b_\ell(\ImgPos)}
  &= \sum_{j,m} w_{j,\ell,m}^2 \, \Var\Brace{y_{j,\ell,m}} \,
                h_{j,\ell,m}^2(\ImgPos) \notag \\
  &= \sum_{j,m} w_{j,\ell,m} \, h_{j,\ell,m}^2(\ImgPos) \notag \\
  &= a_\ell(\ImgPos) \, .
\end{align*}
Using this result, the MLE of the planet SED is given by:
\begin{align}
  \Var\Brace{\estim{\PlanetSED}_\ell(\ImgPos)}
  &= \frac{\Var\Brace{b_\ell(\ImgPos)}}{a_\ell^2(\ImgPos)} \notag\\
  &= \frac{1}{a_\ell(\ImgPos)} \, .
  \label{eq:var-MLE-SED}
\end{align}
The cologarithm of the GLR can be finally put in the form:
\begin{equation}
  \Lambda(\ImgPos) =
  \frac{1}{2} \, \sum_{\ell}
  \frac{\estim{\PlanetSED}_\ell^ 2(\ImgPos)}
       {\Var\Brace{\estim{\PlanetSED}_\ell(\ImgPos)}} \, ,
  \label{eq:colog-GLR-3}
\end{equation}
which is a sum of the squared signal to noise ratio (SNR) of the maximum
likelihood estimator of the planet flux in each spectral channel.  This is a
generalization of a property demonstrated by
\citet{Mugnier_et_al-2009-exoplanet_detection} in the case of multi-frame data.
We therefore introduce:
\begin{equation}
  \mathrm{SNR}(\ImgPos) = \sqrt{2\,\Lambda(\ImgPos)} = \sqrt{
  \sum_{\ell} \frac{b_\ell^2(\ImgPos)}{a_\ell(\ImgPos)}
  } \, .
  \label{eq:detection-SNR}
\end{equation}
which can be thought of as a \emph{detection SNR} accounting for all the spectral
channels.

\citet{Thiebaut_Mugnier-2006-nulling} have shown that accounting for additional
constraints, notably the positivity and regularity of the SED, can greatly
enhance the detection of faint sources.  In our case, it is trivial to find the
maximum likelihood estimator (MLE) of the planet SED subject to the constraint
that it must be nonnegative:
\begin{align}
   \estim{\PlanetSED}_\ell^{+}(\ImgPos)
   &= \argmin_{\PlanetSED_\ell \ge 0} \Brace*{
      b_\ell(\ImgPos) \, \PlanetSED_\ell
      - \frac{1}{2} \, a_\ell(\ImgPos) \, \PlanetSED_\ell^2
    } \notag \\
   &= \frac{\max\Brace{b_\ell(\ImgPos),0}}{a_\ell(\ImgPos)} \, ,
   \label{eq:MLE-nonnegative-planet-SED}
\end{align}
which results from observing that $a_\ell(\ImgPos) > 0$ and is not more
difficult to compute than the unconstrained estimator
$\estim{\PlanetSED}_\ell(\ImgPos)$ in Eq.~(\ref{eq:MLE-planet-SED}).

Combining multi-spectral and multi-temporal data to perform planet detection
has already been proposed by \citet{Thiebaut_Mugnier-2006-nulling} for the
\noun{Darwin} mission.  However, as noted by \citet{Denis_Thiebaut-2015-Lyon}
and in our specific case, it turns out that computing $\Lambda(\ImgPos)$ or
$\mathrm{SNR}(\ImgPos)$ for any assumed planet position $\ImgPos$ on an evenly
spaced grid of positions $\Brace{\ImgPos_i}_{i=1,\ldots}$ can be done in a very
economic way by means of fast Fourier transforms (FFT).


\subsection{Application to simulated data}
\label{sec:simulated-data}

\begin{figure}
  \centering\includegraphics[scale=0.5]{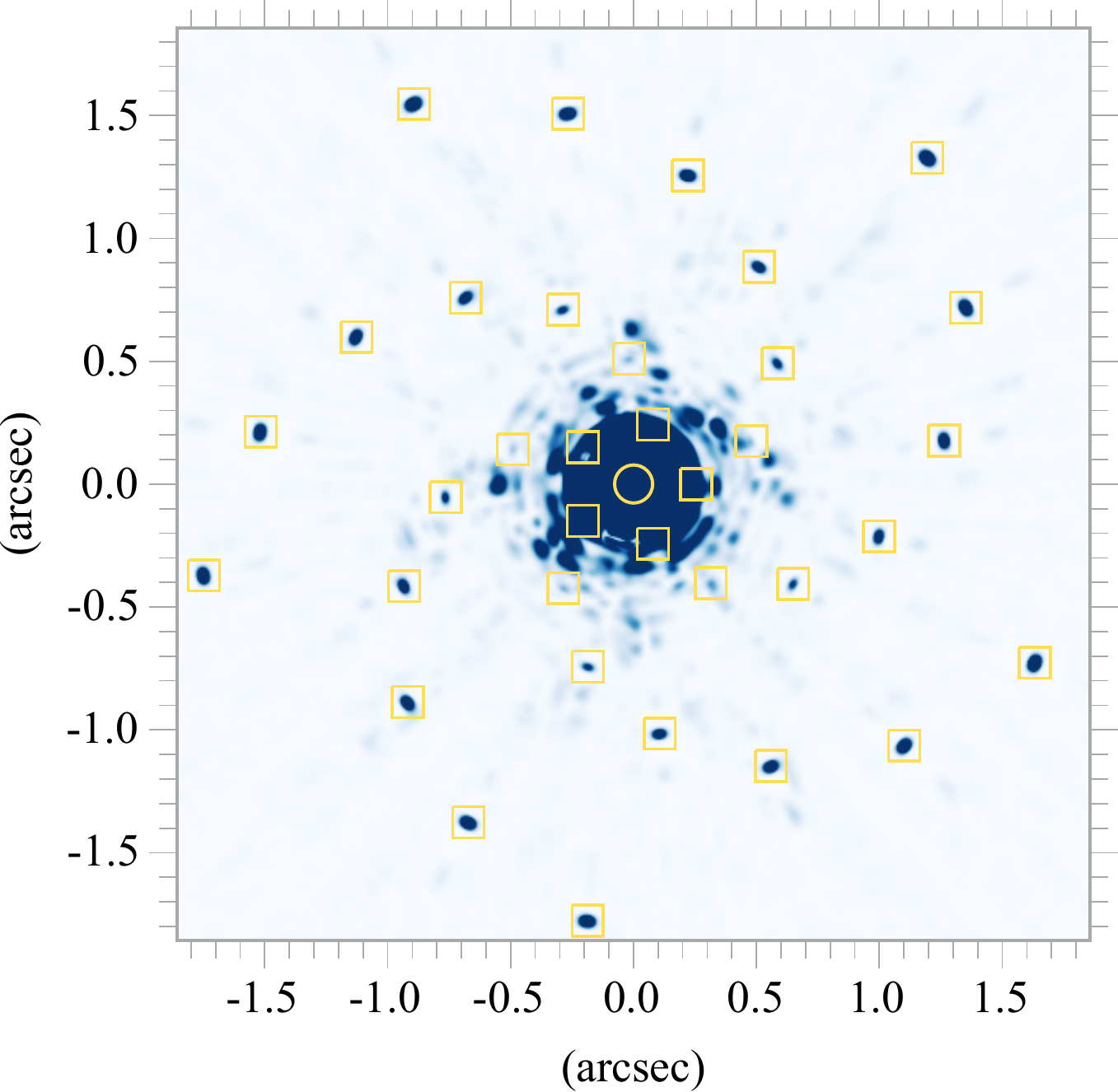}

  \caption{Detection map for planets with a $10^{-6}$ contrast, removal of
  speckles by \PeX model with 3 modes and accounting for the positivity.
  The yellow boxes indicate the positions of planets in the simulation and
  the circle represents the coronograph mask.} \label{fig:detect-maps}
\end{figure}

Figure~\ref{fig:detect-maps} shows a map of the detection SNR as defined in
Eq.~(\ref{eq:detection-SNR}).  To produce this map, we added planets along
spiral tracks to the same simulated speckle patterns as used in the previous
section.  We then applied the proposed \PeX method to estimate and remove the
speckles (as if no planets were present) and compute the detection SNR map for
all the 21 spectral channels.  This map shows that, except under the
coronographic mask and near its edges, our method is able to detect all the
planets which all have the same contrast of $10^{-6}$ with respect to the host
star.  Effectively achieving such a contrast from a single exposure (\ie,
without ADI) is very promising.  We note that a contrast which is constant with
wavelength implies that the planet's SED is the same as that of the host star
(something which may occur for planets with a high albedo) and is the most
unfavorable situation for the detection.  With such a contrast, the planets
cannot be seen without processing the original images which look exactly like
the ones shown in Fig.~\ref{fig:images}.

\subsection{Application to real data}
\label{se:real-data}

\begin{figure}
  \centering \includegraphics[scale=0.43]{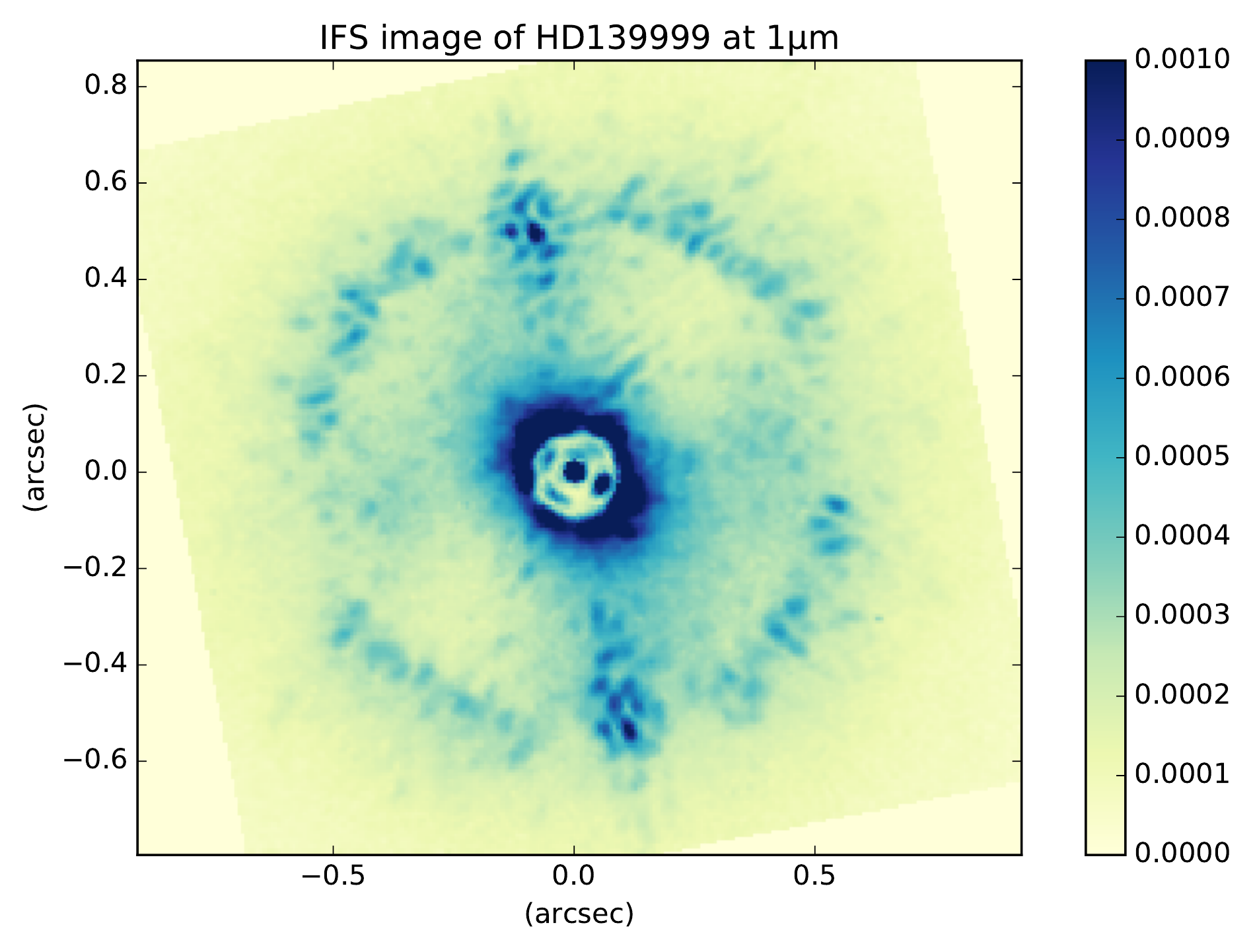}

  \caption{\label{fig:HD139999-image}Real IFS image of HD139999 at $1\,\micron$
   with additional fake planets (which cannot be seen in this image).}
\end{figure}

We also consider applying PeX to real \noun{Sphere} IFS data (\citet{Claudi_SPHERE_IFS}) of the star HD139999A
 to which we added 25 fake planets with a
contrast of $3\times10^{-5}$.  The IFS image at $1\,\micron$ is shown by
Fig.~\ref{fig:HD139999-image}.  As we wanted to demonstrate the ability of our
approach to exploit the chromatism of the speckles, we process all available
spectral channels\footnote{39 spectral channels from $\lambda =
957.5\,\mathrm{nm}$ to $1\,635.8\,\mathrm{nm}$} but only a \emph{single
exposure}.  Note that the previously reported companion, HD139999Ab, which is at about
$840\,\mathrm{mas}$ of HD139999A, is not in the field of the IFS in the
considered exposure \citep{Wagner_et_al-2016-HD139999}.

Assuming a single planet at position $\ImgPos$ and using the same notation as
in Eq.~(\ref{eq:PeX-model2}) and Eq.~(\ref{eq:residuals-model}), the model of
the measured data value in the $j$-th pixel of the $\ell$-th spectral image
writes:
\begin{equation}
  d_{j,\ell} = \underbrace{
    \FirstTermSED_\ell \,
    \sum_{i} R_{i,j,\ell}\,
    \sum_{k=1}^{K} \gamma_\ell^{k - 1} \, \PSFmode_{k,i}
  }_{\displaystyle u_{j,\ell}}
  + \PlanetSED_{\ell} \, h_{j,\ell}(\ImgPos)
  + \varepsilon_{j,\ell} \, ,
 \label{eq:data-model}
\end{equation}
where we drop the exposure index $m$ to simplify the notation and introduce
the linear operator $\M{R}_{\ell}$ to interpolate\footnote{we used Catmull-Rom
bi-cubic interpolation} the diffraction based model of the speckles (expressed
in the reference coordinate system) at the positions of the pixels in the
$\ell$-th spectral image.  Interpolating the model rather than the data avoids
introducing more correlations in the data.  The pixel size of \noun{Sphere} IFS images is
$7.46\,\mathrm{mas}$ and we choose to sample the on-axis PSF modes $\V{p}$ with
an equivalent pixel size of $10\,\mathrm{mas}$ at the reference wavelength
($\lambda^\Tag{ref} = 1\,\micron$).  This sampling size was found to be a good
compromise between spatial smoothness of the speckle model and ability to fit
the finest details.

The unknowns are $\V{q}$, $\V{p}$, $\V{f}$ and $\ImgPos$.  To follow the
procedure described in Section~\ref{sec:detection-criterion}, we first fit
the speckle parameters ($\V{q}$ and $\V{p}$) on the IFS data (assuming
$\V{f} = 0$) and then run the detection tests on the residual
multi-spectral images.  As we already mentioned, fitting our model of the
stellar speckles is difficult because the model is bilinear.  Assuming a
Gaussian distribution of the noise, we solve this problem by a hierarchical
approach which consists in solving:
\begin{equation}
  \estim{\V p} = \argmin_{\V p}\Brace*{\min_{\V q}
  \Norm{\V d -\V u(\V p, \V q)}^2_{\M W}} \, ,
  \label{eq:hierarchical-optimization}
\end{equation}
where $\Norm{\V d -\V u(\V p, \V q)}^2_{\M W}$ is the $\chi^2$ of the data $\V
d$ given the model $\V u(\V p, \V q)$ of the speckles introduced in
Eq.~(\ref{eq:data-model}).  Here $\Norm{\V y}^2_{\M W} = \V y\T\cdot\M W\cdot\V
y$ is a weighted quadratic norm and the weight $\M W$ is the inverse of the
noise covariance.  Since the $\chi^2$ is quadratic in $\V q$,
the innermost minimization in Eq.~(\ref{eq:hierarchical-optimization}) is
straightforward.  To carry the outermost minimization, we used a non-linear
quasi-Newton method \citep{Nocedal-1980-vmlm} to optimize over the parameters $\V
p$.  We found that, in practice, this hierarchical optimization strategy was
very effective.

\begin{figure}
  \centering
  \includegraphics[scale=0.43]{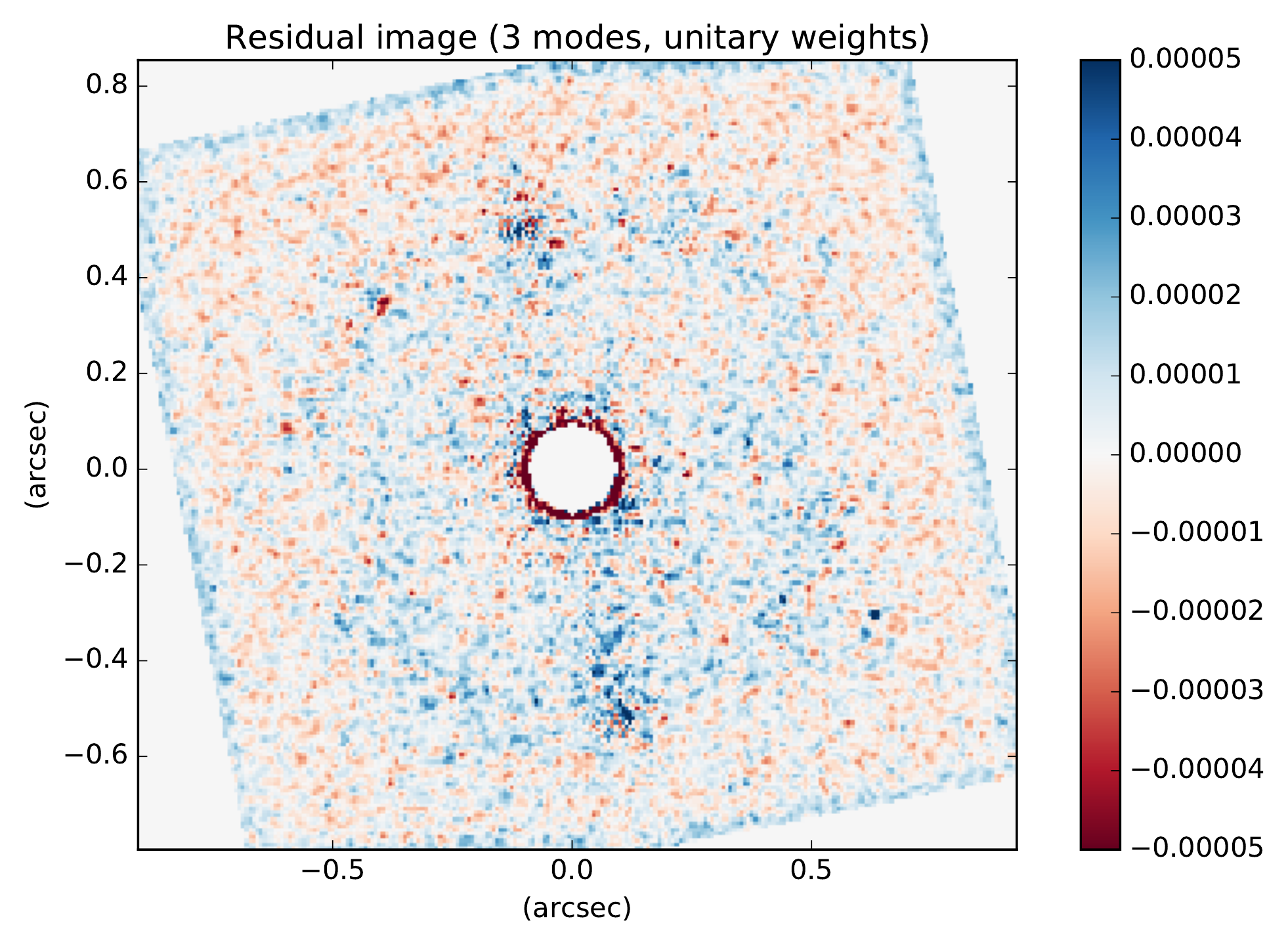}

  \includegraphics[scale=0.43]{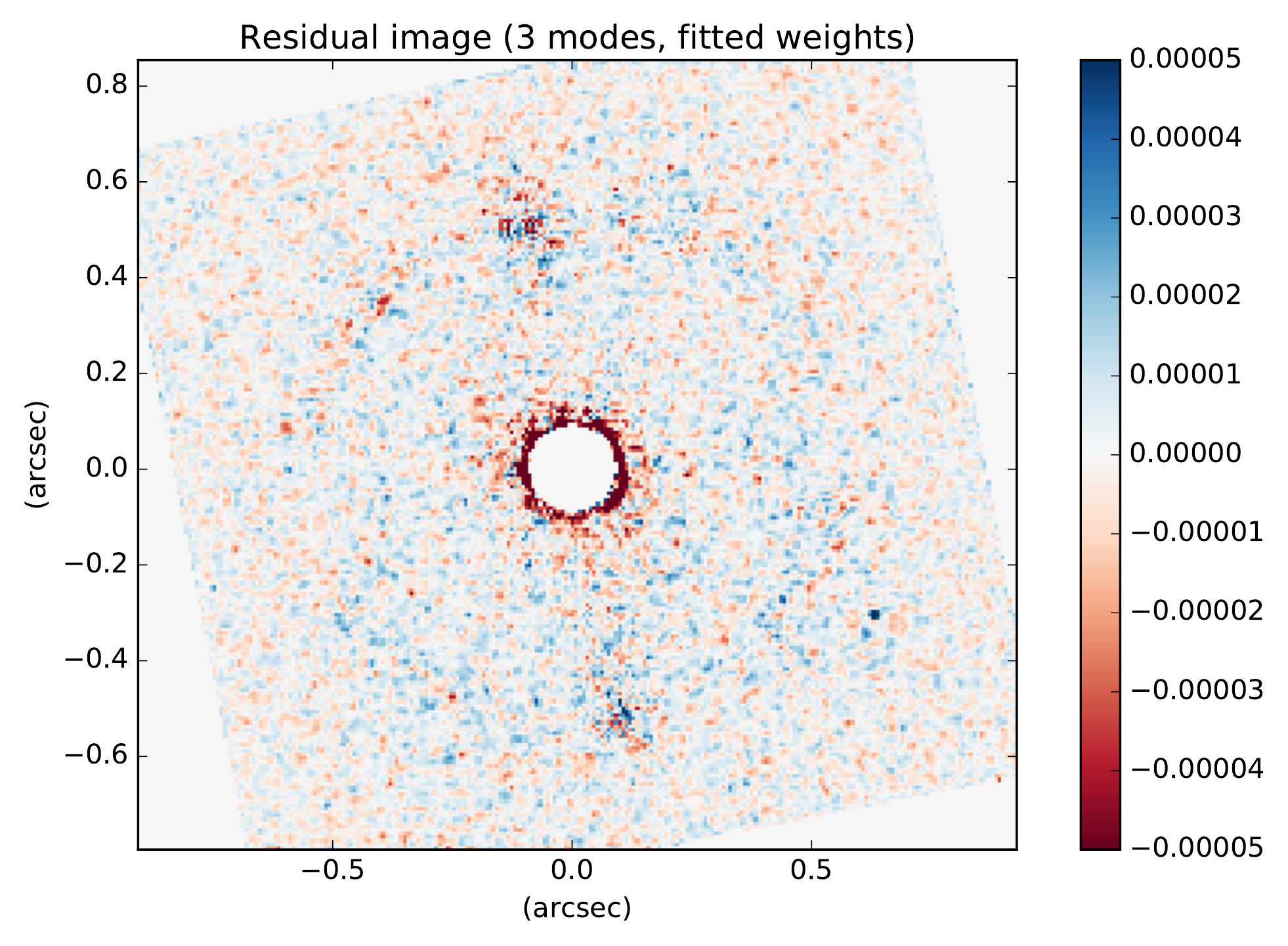}

  \caption{\label{fig:HD139999-residuals}Residual images after removing the
  PeX model of the speckles.  Here only 3 modes have been used to model the
  speckles. The top image shows the residuals assuming uniform weights
  while the weights computed from the simple model of the noise variance
  have been used for the bottom image.}
\end{figure}

Compared to the simulations in the previous Section, the data are corrupted by
noise and a correct estimation of the statistical weights $\M W$ is very
important.  As a first simplification, we assumed independent data and thus a
diagonal weighting operator $\M W$ whose components can be computed from the
noise variance as in Eq.~(\ref{eq:statistical-weights}).  Since no estimation
of the noise variance is provided with \noun{Sphere} IFS data, we estimated this
variance by assuming the following simple model
\citep{Foi_et_al-2008-Poissonian_Gaussian_noise_model_fitting}:
\begin{equation}
  \Var(\varepsilon_{j,\ell}) = \alpha_{\ell}\,\Expect\{d_{j,\ell}\}
  + \beta_{\ell} \, ,
  \label{eq:variance-model}
\end{equation}
where $\alpha_{\ell} \ge 0$ and $\beta_{\ell} > 0$ are two unknown parameters
which we assume to be the same for all the pixels of a given spectral channel
$\ell$.  The term $\alpha_{\ell}\,\Expect\{d_{j,\ell}\}$ is the variance due to
the photon noise while $\beta_{\ell}$ is the variance of the detector noise.
The parameters $\alpha_{\ell}$ depend on the quantum efficiency and on the gain
of the detector.  We assumed that $\Expect\{\V d\} \approx \V u(\V p, \V q)$ and derive the noise model parameters from their maximum likelihood values:
\begin{equation}
  \Paren{\estim{\alpha}_\ell, \estim{\beta}_{\ell}} =
    \argmin_{\alpha,\beta} \sum_j \Brace*{
      \frac{(d_{j,\ell} - u_{j,\ell})^2}{ \alpha\,u_{j,\ell} + \beta}
      + \log(\alpha\,u_{j,\ell} + \beta)
    } \, ,
\end{equation}
where the first term in the sum is the $\chi^2$ of the data in the $j$-th pixel
of the $\ell$-th spectral channel while the logarithm term is due to the
normalization of the assumed Gaussian distribution.  We use Powell's BOBYQA
algorithm \citep{Powell-2009-BOBYQA} to solve the above problem.  As the noise
parameters depend on the model $\V u(\V p, \V q)$ of the speckles which
themselves depend on the weights and hence on the noise model parameters, we
apply the following alternating procedure: (i) assume uniform weights (ii) fit
the model of the speckles (iii) fit the noise model parameters (iv) update
the weights and repeat starting at step (ii) until convergence.  In practice, this
procedure is stable and about 3 to 5 iterations are sufficient.  The comparison
of the residuals shown in Fig.~\ref{fig:HD139999-residuals} demonstrates that
non-uniform weights fitted by the proposed alternating method yields smaller
and less structured residuals.

\begin{figure}
\centering
\includegraphics[scale=0.43]{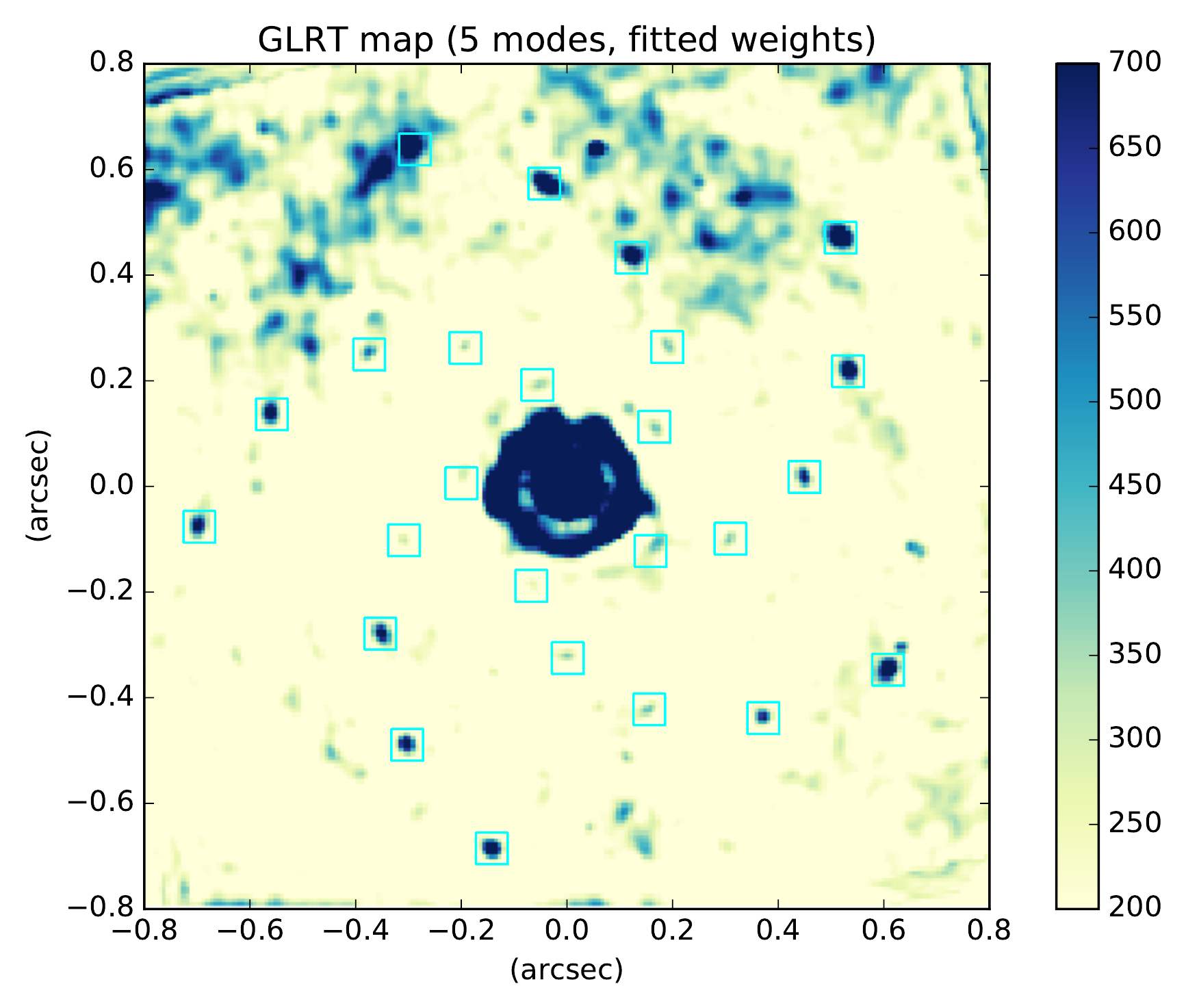}
\caption{\label{fig:HD139999-glrt}Map of the GLRT criterion. 5 modes have been used to model the speckles.  The light blue boxes indicate the locations of the fake planets, all have a contrast of $3\times10^{-5}$.}
\end{figure}

\begin{figure}
  \centering
  \includegraphics[scale=0.43]{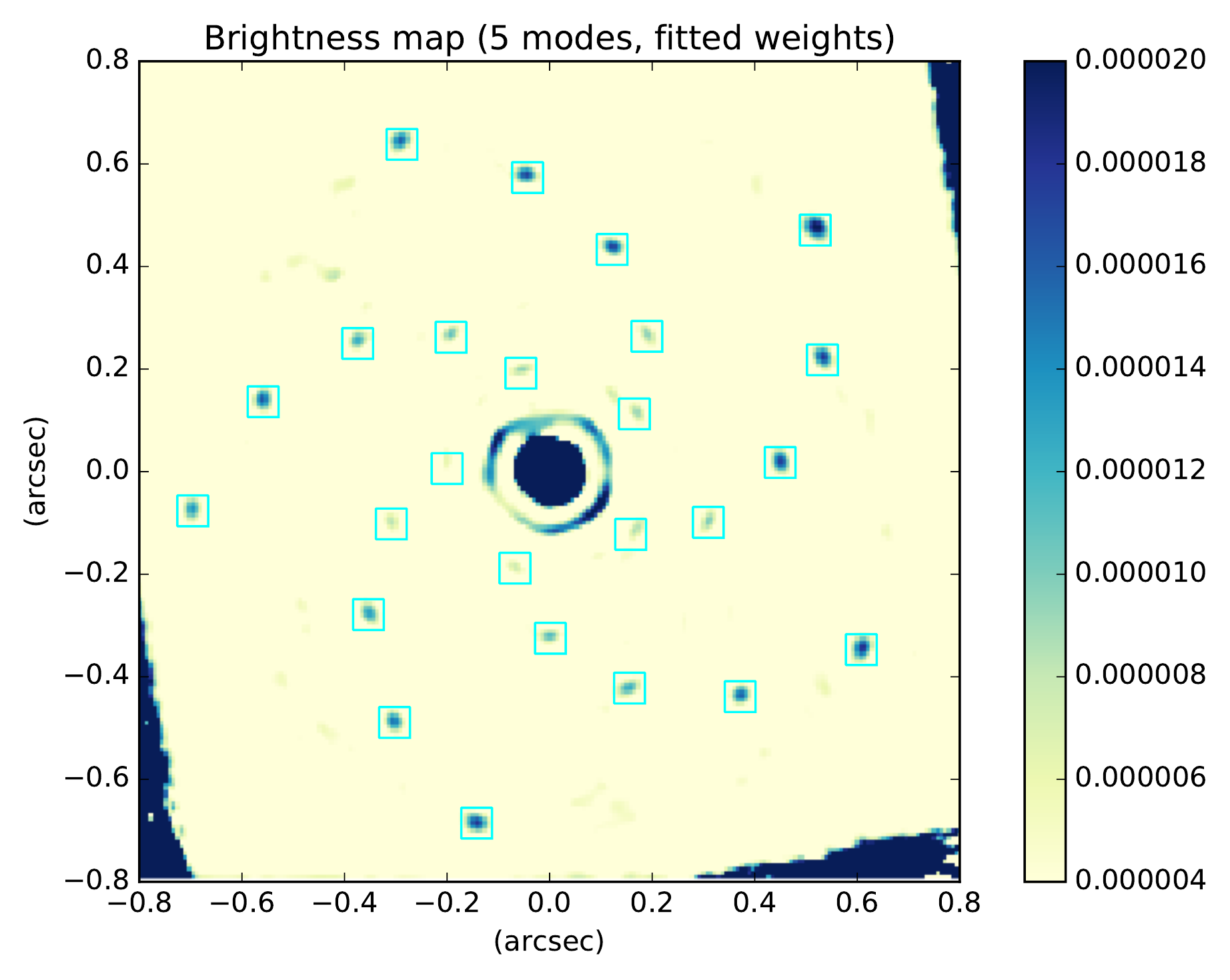}

  \caption{\label{fig:HD139999-brightness}Map of the most likely brightness of
  point-like sources. 5 modes have been used to model the speckles.  The light
  blue boxes indicate the locations of the fake planets, all have a contrast of
  $3\times10^{-5}$.}
\end{figure}

Finally we apply the detection tests described in
Section~\ref{sec:detection-criterion} to the residuals $\V d - \V u(\estim{\V
p}, \estim{\V q})$, Figure~\ref{fig:HD139999-glrt} shows the generalized
likelihood ratio (GLR) $\Lambda(\V r)$ for the data in the considered exposure.
All the fake planets have a GLR which is a local maximum but not all can be
detected without false alarms as there are several other positions where the
criterion is higher (for instance, in the upper part of the field of view).
Looking at Equations~(\ref{eq:colog-GLR-2}), (\ref{eq:glr-den}) and
(\ref{eq:glr-num}) it is evident that any errors in the magnitude of the
statistical weights (which do not appear with the same power in the numerator
and denominator of the GLR) could lead to a grossly wrong GLR.  The assumed
model of the noise variance, in Eq.~(\ref{eq:variance-model}) is too simple, at
least because correlation in the data are ignored.  Indeed, due to the way IFS
multi-spectral images are produced, nearby pixels and spectral channels are
strongly correlated.  To mitigate this issue, we could have compared the GLR to
its mean or median value along circular tracks at the same distance from the
host star.  We however note that the statistical weights appear with the same power in the numerator and
denominator of the planet SED given by Eq.~(\ref{eq:MLE-planet-SED}) and we
therefore expect that errors in the magnitude of the weights somewhat
compensate in the estimated SED even though the estimator is no longer optimal.
We therefore assumed a constant SED for the sought planets (\ie,
$\PlanetSED_\ell = \PlanetSED$, $\forall \ell$) and compute a map of
the \emph{best planet brightness} given its assumed position $\ImgPos$ which is
simply given by:
\begin{align}
   \estim{\PlanetSED}(\ImgPos)
   &= \argmin_{\PlanetSED \ge 0} \sum_{\ell} \Brace*{
      b_\ell(\ImgPos) \, \PlanetSED
      - \frac{1}{2} \, a_\ell(\ImgPos) \, \PlanetSED^2
    } \notag \\
   &= \frac{
     \max\Paren[\big]{0, \sum_{\ell} b_\ell(\ImgPos)}
   }{
     \sum_{\ell} a_\ell(\ImgPos)
   }\, .
   \label{eq:MLE-planet-brightness}
\end{align}
In this map, shown by Fig.~\ref{fig:HD139999-brightness}, all the fake planets can
be clearly seen with perhaps 1 or 2 false alarms.  As a consequence of
estimating the stellar leakage and then the planetary signal (if any), the
estimated planet brightnesses in Fig.~\ref{fig:HD139999-brightness} are always
significantly lower than their true value: between $\PlanetSED \approx
2\times10^{-5}$ for the most remote planets and $\PlanetSED \approx
6\times10^{-6}$ for the ones close to the host star while the truth is
$\PlanetSED = 3\times10^{-5}$.  A joint estimation of all these unknowns given
the data should yield the best results and give an unbiased estimate of the planet SED.

These results on empirical data are very encouraging, notably because they were
obtained with a single IFS exposure.  To improve the detection limit, multiple
exposures could be combined, but it is perhaps more important that the
correlations in the data be taken into account.  Performing a joint estimation
of all the parameters would also be an improvement.

\section{Discussion}

Our aim in this paper is to enhance the removal of residual speckle in
multi-wavelength images in order to improve exoplanet detection limits. In
order to do this we have extended the PSF expansion of
\citet{Perrin_et_al-2003-PSF} to explicitly take into account wavelength
dependence. We show that the PSF may be written as a combination of spatial
modes which spatially scale with the wavelength and which are multiplied by
chromatic factors with a power law dependence on wavelength and mode number.
The exact power-law depends on which mode dominates the residuals, and has to
be estimated from the data.

If the multi-wavelength data is rescaled to a reference wavelength, then the
model is separable and is given by an expansion of spatial modes multiplied by
a wavelength-dependent factor.  We refer to this as the P{\sc e}X model of the
data. The chromatic factor is the product of the SED of the star and a
power-law with an exponent $\beta$.

In order to verify our model we simulated multi-wavelength data (over the H
band) from a coronographic system with characteristics similar to the
\noun{Sphere} exoplanet integral field spectrograph (IFS).  The data cube is
rescaled to $1.65\,\micron$, and SVD used to fit a separable model. In the
first instance the SVD is truncated to fit a single mode and the fit is carried
out in annuli centred on the axis. It is found that outside the coronographic
mask a power law indeed provides an almost perfect fit to the corresponding
chromatic factor. In the absence of aberrations, the fitted power-law exponent
implies that the zeroth order mode dominates, as expected. When aberrations are
added to the simulation (70\,nm rms) the fitted exponent depends on distance
from the axis; but the second order mode dominates in the wings. This is
exactly what is expected from the Perrin et al.\ analysis. We investigated
reducing the speckle residuals by subtracting single modes from the data, using
either the SVD modes or fitting a mode obtained using the \PeX model. When the
\PeX model uses the correct chromatic exponent, the performance is very similar
to SVD, and in fact can reduce the speckles close to the edge of the
coronographic mask.

The SVD fitting is subsequently carried out using multiple modes. It is found
that the speckle suppression becomes insensitive to the exact choice of
spectral exponents when more than 3 or 4 modes are fitted. In the simulated
data, the level of suppression reaches $10^{-7}$ near the edge of the mask
using just 5 modes.

Assuming Gaussian independent noise, which should be suitable for well-cleaned
residuals, we derive the Maximum Likelihood Estimator (MLE) for the planet flux
at each wavelength, and its variance.  The optimal detection criterion then
amounts to finding the planet position which has the maximal detection SNR (\cf
Eq.~(\ref{eq:detection-SNR})). By adding fake planets to the simulated data we
demonstrate detection down to a contrast ratio of $10^{-6}$ from a single
exposure, although some speckle at this level can be seen near the edge of the
coronographic mask. With real IFS data, we were able to achieve
detection of fake planets from a \emph{single exposure} with a contrast of
$3\times10^{-5}$ at $200\,\mathrm{mas}$ from the center.  This limit compares
favorably to other methods but is not as good as with simulated data.  This is
due to the noise in the real data (in our simulations there is no added noise)
and to the assumption that pixels are independent (which is not the case with
IFS multi-spectral images).  By combining independent exposures and exploiting
the apparent motion of the sources in the field of view, we however expect to
improve the contrast limit by a factor roughly equal to the square root of the
number of exposures.

Compared to other techniques for exoplanet detection in multi-spectral data we
believe that our approach offers some important advantages. It is based on a
physical model of the residual PSF, which provides some insight compared to
ad-hoc approaches. It is well suited to simultaneous speckle suppression and
planet detection, which we are developing for a future publication. In the
current application, we fit modes to a data cube made up of the re-scaled and
interpolated narrow-band images. The fitting could be carried out on the
original images by taking the re-scaling into account explicitly, thereby
removing the need for interpolation which can introduce artifacts. In fact, the
inverse approach could be applied to the IFS raw data.

Most current approaches to processing ADI data are empirical and somewhat
ad-hoc. However, some efforts have been made to develop algorithms which are
statistically optimal.  For example,
\citet{Smith_et_al-2009-exoplanet_detection} describe a Maximum Likelihood
approach to jointly estimate the stellar PSF and the planet position and
intensity from the data while \citet{Mugnier_et_al-2009-exoplanet_detection}
describes a Maximum Likelihood approach to detecting planets in ADI images
which have been pair-wise subtracted. The analysis presented here can be used
to process multi-temporal data, taking into account any possible rotation or
other transformation of the data as a function of time, as well as temporal
correlation of the PSF modes. This work is under development and will be
demonstrated in a a subsequent paper.

\section*{Acknowledgements}

The authors are very grateful to Maud Langlois for useful discussions to
help understand the instrument and for carefully preprocessing the
HD139999 data.

This work has made use of the SPHERE Data Centre, jointly operated by
OSUG/IPAG (Grenoble), PYTHEAS /LAM/CeSAM (Marseille), OCA/Lagrange
(Nice) and Observatoire de Paris/LESIA (Paris).

The research leading to these results has received support from the DETECTION
project funded by the French CNRS (Mission pour l'Interdisciplinarité, DEFI
IMAGIN) and from the \emph{Programme Avenir Lyon Saint-Étienne Projet
Emergent PALSE/2013/26}.

The simulations have been carried out using the \noun{Yorick} language
\citep{Munro-1995-Yorick} while the empirical data has been processed using
the \noun{Julia} language \citep{Bezanson_et_al-2017-Julia}.

\bibliographystyle{abbrvnat}
\bibliography{pex-biblio}

\end{document}